\documentclass[longauth]{aa}  

\usepackage{graphicx}
\usepackage{txfonts}
\usepackage{multirow}
\usepackage{amsmath}
\usepackage{url}
\usepackage{xcolor}
\begin{document}

   \title{CHEOPS observations of the HD\,108236 planetary system: A fifth planet, improved ephemerides, and planetary radii}

   \titlerunning{CHEOPS observations of the HD\,108236 planetary system}
   \authorrunning{Bonfanti et al.}

   \author{A.~Bonfanti\inst{1}
\and
L.~Delrez\inst{2,3,4}
\and M.J.~Hooton\inst{5}
\and T.G.~Wilson\inst{6}
\and L.~Fossati\inst{1}
\and Y.~Alibert\inst{5}
\and S.~Hoyer\inst{7}
\and A.J.~Mustill\inst{8}
\and H.P.~Osborn\inst{9,10}
\and V.~Adibekyan\inst{11}
% ...
}

   \institute{Space Research Institute, Austrian Academy of Sciences, 
              Schmiedlstrasse 6, A-8042 Graz, Austria\\
              \email{andrea.bonfanti@oeaw.ac.at}
\and
Space sciences, Technologies and Astrophysics Research (STAR) Institute, Université de Liège, Allée du six Août 19C, 4000 Liège, Belgium
\and
Astrobiology Research Unit, Université de Liège, Allée du six Août 19C, 4000 Liège, Belgium
\and
Observatoire de Genève, Université de Genève, Chemin des Maillettes 51, 1290 Sauverny, Switzerland
\and
Physikalisches Institut, University of Bern, Gesellsschaftstrasse 6, 3012 Bern, Switzerland
\and
School of Physics and Astronomy, Physical Science Building, North Haugh, St Andrews, United Kingdom
\and
Aix Marseille Univ, CNRS, CNES, LAM, Marseille, France
\and
Lund Observatory, Dept. of Astronomy and Theoreical Physics, Lund University, Box 43, 22100 Lund, Sweden
\and
NCCR/PlanetS, Centre for Space \& Habitability, University of Bern, Bern, Switzerland
\and
Department of Physics and Kavli Institute for Astrophysics and Space Research, Massachusetts Institute of Technology, Cambridge, MA 02139, USA
\and
Instituto de Astrof\'isica e Ci\^encias do Espa\c{c}o, Universidade do Porto, CAUP, Rua das Estrelas, 4150-762 Porto, Portugal
% ...
}

   \date{}

 
  \abstract
   {The detection of a super-Earth and three mini-Neptunes transiting the bright ($V$\,=\,9.2\,mag) star HD\,108236 (also known as TOI-1233) was recently reported on the basis of {\it TESS} and ground-based light curves.}
   {We perform a first characterisation of the HD\,108236 planetary system through high-precision {\it CHEOPS} photometry and improve the transit ephemerides and system parameters.}
   {We characterise the host star through spectroscopic analysis and derive the radius with the infrared flux method. We constrain the stellar mass and age by combining the results obtained from two sets of stellar evolutionary tracks. We analyse the available {\it TESS} light curves and one {\it CHEOPS} transit light curve for each known planet in the system.}
   {We find that HD\,108236 is a Sun-like star with $R_{\star}=0.877\pm0.008\, R_{\odot}$, $M_{\star}=0.869^{+0.050}_{-0.048}\, M_{\odot}$, and an age of $6.7_{-5.1}^{+4.0}$ Gyr. We report the serendipitous detection of an additional planet, HD\,108236\,f, in one of the {\it CHEOPS} light curves. For this planet, the combined analysis of the {\it TESS} and {\it CHEOPS} light curves leads to a tentative orbital period of about 29.5\,days. From the light curve analysis, we obtain radii of $1.615\pm0.051$, $2.071\pm0.052$, $2.539_{-0.065}^{+0.062}$, $3.083\pm0.052$, and $2.017_{-0.057}^{+0.052}$ $R_{\oplus}$ for planets HD\,108236\,b to HD\,108236\,f, respectively. These values are in agreement with previous {\it TESS}-based estimates, but with an improved precision of about a factor of two. We perform a stability analysis of the system, concluding that the planetary orbits most likely have eccentricities smaller than 0.1. We also employ a planetary atmospheric evolution framework to constrain the masses of the five planets, concluding that HD\,108236\,b and HD\,108236\,c should have an Earth-like density, while the outer planets should host a low mean molecular weight envelope.}
   {The detection of the fifth planet makes HD\,108236 the third system brighter than $V$\,=\,10\,mag to host more than four transiting planets. The longer time span enables us to significantly improve the orbital ephemerides such that the uncertainty on the transit times will be of the order of minutes for the years to come. A comparison of the results obtained from the {\it TESS} and {\it CHEOPS} light curves indicates that for a $V$\,$\sim$\,9\,mag solar-like star and a transit signal of $\sim$500\,ppm, one {\it CHEOPS} transit light curve ensures the same level of photometric precision as eight {\it TESS} transits combined, although this conclusion depends on the length and position of the gaps in the light curve.}


   \maketitle
%

\section{Introduction}
Transiting exoplanets provide the unique opportunity to thoroughly characterise planetary systems, from atmospheres to orbital dynamics, and transiting multi-planet systems play a special role. Multi-planet systems enable one, for example, to identify the presence of orbital resonances among the detected planets in a system, giving the possibility to use transit timing variations (TTVs) to measure planetary masses and/or detect other planets in the system \citep[e.g.][]{miralda2002,holman2005,agol2005}.

There are many additional reasons why multi-planet systems are of particular interest. The existence of multiple planets that formed in the same disk places stronger constraints on formation models relative to planets in isolation, motivating the quantification of orbital spacings, correlations and differences within a system \citep{lissauer2011,fabrycky2014,winn2015,weiss2018}, and inspiring novel approaches to classification and statistical description \citep{alibert2019,sandford2019,gilbert2020}. The spacing of planets relative to mean motion resonances provides information about planetary migration during formation, as well as later tidal effects on orbits \citep{delisle2012,izidoro2017}. A system's multiplicity is affected after formation by long-term orbital dynamics, whether driven internally or as a result of more distant undetected planetary perturbers \citep{pu2015,mustill2017,he2020}. Changes to orbits can even affect the climate of planets \citep{spiegel2010}. Given that our own Solar System contains multiple planets, this all helps us to understand points of similarity and divergence between our system and others.

Multi-planet systems also offer the opportunity to study the correlation between the composition (bulk and/or atmospheric) of planets and their periods or equilibrium temperatures, particularly when both planetary masses and radii have been measured. This correlation is a powerful constraint on planet formation and composition models as the number of degrees of freedom one can play with in models is reduced by the fact that all planets formed in the same protoplanetary disk. However, observing such a correlation requires precise transit measurements and dynamical analyses (to assess mass values via TTVs or radial velocity follow-up), which in turn can be more easily done once precise ephemerides of the different planets in the system are known. 

Finally, multi-planet systems are ideal laboratories for studying the evolution of planetary atmospheres. This process is controlled by the host star's evolution (i.e. evolution of the stellar radius, mass, and high-energy radiation), by the physical characteristics of each planet (e.g. planetary mass, radius, and initial atmospheric mass fraction and composition), and by the orbital evolution of each planet. Within multi-planet systems, each planet evolved in its own way as a result of its specific planetary and orbital characteristics, but the range of possible evolutionary paths is limited by the fact that all planets in the system orbit the same star. This enables one not only to constrain the evolution history of the planets, but also aspects of the host star that would be unattainable otherwise, such as the evolution of the stellar rotation rate \citep[e.g.][]{kubyshkina19a,kubyshkina19b,owen2020a}.

The majority of transiting multi-planet systems known to date were detected by the {\it Kepler} and {\it K2} missions \citep[e.g.][]{coughlin2016,mayo2018}. Among these, about 60 systems host four or more transiting planets, but only two have a host star brighter than $V$\,=\,10\,mag\footnote{From {\tt https://exoplanetarchive.ipac.caltech.edu/}} (Kepler-444: \citeauthor{campante2015} \citeyear{campante2015}; HIP\,41378: \citeauthor{vanderburg2016} \citeyear{vanderburg2016}). The launches of the {\it TESS} \citep[Transiting Exoplanet Survey Satellite;][]{ricker2015} and {\it CHEOPS} \citep[CHaracterising ExOPlanets Satellite;][]{benz20} satellites have shifted the focus of the detection and characterisation of multi-planet systems towards brighter stars. While {\it TESS}, similarly to {\it Kepler} and {\it K2}, has a wide field of view (FoV) that is optimised for the detection of a large number of transiting planets, {\it CHEOPS} is a targeted mission, observing one system at a time to perform a precise characterisation.

As of 19 November 2020, {\it TESS} had discovered 82 confirmed planets and $\sim$\,60\% of them belong to multi-planet systems. A non-exhaustive list of the multi-planet systems discovered by {\it TESS} includes HD\,15337 \citep{gandolfi19}, TOI-125 \citep{quinn19}, HD\,21749 \citep{dragomir19}, HR\,858 \citep{vanderburg19}, LP\,791-18 \citep{crossfield19}, L98-59 \citep{kostov19}, TOI-421 \citep{carleo20}, HD\,63433 \citep{mann20}, and TOI-700 \citep{gilbertEA20}. Following the completion of its prime mission on 5 July 2020, {\it TESS} was extended for a further 27 months. This will not only allow us to re-observe many of the targets already studied during the prime mission to better characterise them, but also to observe additional stars for the first time.

\citet[][D20 hereafter]{Daylan2020} announced the detection with {\it TESS} of four transiting planets orbiting the bright ($V$\,=\,9.2\,mag) solar-like star HD\,108236. The four planets have periods of about 3.8, 6.2, 14.2, and 19.6 days. The radius of the innermost planet ($\approx$1.6\,$R_{\oplus}$) suggests that this is possibly a rocky super-Earth, while the larger radii of the three outer planets ($\approx$2.1, 2.7, and 3.1\,$R_{\oplus}$) indicate that they may still host a lightweight gaseous envelope \citep{fulton2017,james17,jin18}. From the {\it TESS} measurements, it follows that the inner planet lies inside the radius gap \citep{fulton2017}, while the three larger outer planets are located around the peak comprising planets with a gaseous envelope, hence making this system of particular interest for atmospheric evolution studies.

D20 performed orbital dynamic simulations that showed that the system is stable, though a significant exchange of angular momentum among the planets in the system likely occurred. Furthermore, on the basis of these simulations, D20 suggested the possible presence of a fifth planet in the system with a period of 10.9\,days. However, a dedicated analysis of the {\it TESS} light curve (LC) did not give definitive proof. Finally, the bright host star makes the HD\,108236 system a primary target for planetary mass measurements through radial velocities (RVs) and for constraining the atmospheric properties of multi-planet systems (D20).

We report here the results obtained from {\it CHEOPS} high-precision photometric observations of one transit of each detected planet composing the HD\,108236 system, taken almost one year after the {\it TESS} observations. The main goals of the observations presented here were to secure the ephemerides of all detected planets, to employ the exquisite quality of {\it CHEOPS} photometry to provide a first refinement of the system's main properties, and to confirm or disprove the presence of the putative fifth planet at the approximately 10.9\,days indicated by D20. 

This paper is organised as follows. Section \ref{sec:star} presents the host star properties, and Section \ref{sec:observations} describes the {\it CHEOPS} and {\it TESS} LCs. The data analysis is presented in Section \ref{sec:analysis}, and results are reported in Section \ref{sec:results}. Section \ref{sec:conclusions} summarises the work and presents our conclusions.

\section{Host star properties}\label{sec:star}
HD\,108236 is a bright Sun-like star (spectral type G3V) that is also known as HIP 60689, TOI-1233, and Gaia DR2 6125644402384918784.
Between 13 December 2019 and 23 January 2020 (UT) we acquired 13 high-resolution spectra ($R$\,=\,115\,000) of HD\,108236 (programme ID 1102.C-0923, PI: Gandolfi) using the High Accuracy Radial velocity Planet Searcher \citep[HARPS,][]{Mayor2003} spectrograph mounted at the ESO-3.6\,m telescope of La Silla Observatory, Chile. We set the exposure time to  T$_\mathrm{exp}$\,=\,1100--1500 s depending on the sky and seeing conditions, which led to an average signal to noise ratio (S/N) per pixel of $\sim$100 at 550\,nm. We used the co-added HARPS spectrum -- which has a consequent S/N\,$\sim$\,360 per pixel -- to derive the fundamental photospheric parameters of the star, namely, the effective temperature $T_{\mathrm{eff}}$, surface gravity $\log{g}$, and metal content [Fe/H].
We obtained $T_{\mathrm{eff}}=5660\pm61$\,K, $\log{g}=4.49\pm0.11$, and [Fe/H]$=-0.28\pm0.04$\,dex, from spectral analysis, which made use of the ARES+MOOG tools \citep{sousa14}. In short, we measured the equivalent widths of iron lines using the ARES code\footnote{The latest version of the ARES code (ARES v2) can be downloaded at {\tt http://www.astro.up.pt/$\sim$sousasag/ares}.} \citep{Sousa-07, Sousa-15} on the combined HARPS spectrum. In this step, we used the list of iron lines presented in \citet[][]{Sousa-08}. The best fitting atmospheric parameters were obtained looking for convergence of both ionisation and excitation equilibria. For this step, we made use of a grid of Kurucz model atmospheres \citep{Kurucz-93} and the radiative transfer code MOOG \citep{Sneden-73}. We also analysed the same spectra using the 'spectroscopy made easy' (SME) code \citep{piskunov17}, which uses a different method and a different grid of models \citep[\textsc{atlas12},][]{kurucz13} achieving results well within 1 sigma of those obtained employing ARES+MOOG.

It has been suggested that individual abundances of heavy elements and specific elemental ratios end up controlling the structure and composition of the planets \citep[e.g.][]{bond10,thiabaud15,unterborn16,santos15}. In particular, Mg/Si and Fe/Si mineralogical ratios were proposed as probes to constrain the internal structure of terrestrial planets \citep[e.g.][]{dorn15}. Therefore, we specifically derived the abundances of Mg and Si using the same tools and models as for the atmospheric parameter determination, as well as using the classical curve-of-growth analysis method assuming local thermodynamic equilibrium.
For deriving the abundances, we closely followed the methods described in \citet{Adibekyan-12} and \citet{Adibekyan-15}. The solar reference Mg and Si abundances are taken from \citep{asplund09}. The Mg/Si and Fe/Si abundance ratios were calculated as
\begin{equation} 
 A/B = N_{A}/N_{B} = 10^{\log \epsilon(A)}/10^{\log \epsilon(B)}\,,
 \label{eq:abund}
\end{equation}
where $N_A$ and $N_B$ represent the number of atoms of elements A and B, respectively, scaled assuming a hydrogen content of $10^{12}$ atoms, while $\log \epsilon(A)$ and $\log \epsilon(B)$ are the respective absolute elemental abundances (total number of atoms) expressed in logarithmic scale.

We employed the infrared flux method (IRFM; \citealt{Blackwell1977}) to calculate the radius of the host star through the determination of the stellar angular diameter $\theta$ and effective temperature using known relationships between these properties, optical and infrared broadband fluxes, and synthetic photometry obtained from stellar atmospheric models \citep{Castelli2003} over various standard bandpasses. The {\it Gaia} G, G$_{\rm BP}$, and G$_{\rm RP}$, the 2MASS J, H, and K, and the {\it WISE} W1 and W2 fluxes and relative uncertainties were retrieved from the most recent data releases \citep[][respectively]{GaiaCollaboration2018,Skrutskie2006,Wright2010}. We applied a Markov chain Monte Carlo (MCMC) approach, setting priors on the stellar parameters taken from the spectroscopic analysis detailed above. Within this framework, accounting for the reddening $E(B-V)$, we compared the observed photometry with the synthetic one obtained from convolving stellar synthetic spectral energy distributions from the \textsc{atlas} Catalogues \citep{Castelli2003} with the throughput of the considered photometric bands. From this analysis, we determined the stellar radius and $E(B-V)$ to be $R_{\mathrm{IRFM},\star}=0.877\pm0.008\, R_{\odot}$ and $E(B-V)_{\rm IRFM}=0.12\pm0.09$, respectively. These values are in agreement with those provided in the literature (D20), but have a precision on the stellar radius of twice that previously reported.

The stellar mass $M_{\star}$ and age $t_{\star}$ were inferred from evolutionary models. To obtain more robust results, we considered two different sets of tracks and isochrones: one set generated from the PARSEC\footnote{Padova and Trieste Stellar Evolutionary Code\\ {\tt http://stev.oapd.inaf.it/cgi-bin/cmd}.} v1.2S code \citep{marigo17} and another with the CLES code \citep[Code Liègeois d'Évolution Stellaire;][]{scuflaire08}. The two models differ for example in terms of solar mixture, helium-to-metal enrichment ratio, adopted reaction rates, and opacity and overshooting treatment. The PARSEC models adopt the solar-scaled composition given by \citet{caffau11}, while the CLES models consider that given by \citet{asplund09}. In PARSEC, the helium content $Y$ is assumed to increase with $Z$ according to a linear relation of the form $Y=\frac{\Delta Y}{\Delta Z}Z + Y_p$, where $\frac{\Delta Y}{\Delta Z}=1.78$ has been inferred from solar calibration and $Y_p=0.2485$ is the primordial helium abundance \citep{komatsu11}; instead, in CLES the helium content may vary regardless of the heavy elements abundances, or be fixed following a metal enrichment linear law as above. The PARSEC models have been computed considering the nuclear reaction rates given by the JINA REACLIB database \citep{cyburt10}, while the CLES models consider the compilation from \citet{Adelberger11}. For hot stars, PARSEC models use the opacities based on the Opacity Project \citep[OP,][]{seaton05}, while CLES models consider the OPAL tables of opacities \citep{Iglesias96}; in the low-temperature regime, PARSEC complements its opacity database with {\AE}SOPUS opacities \citep{marigo09}, while CLES with the opacities taken from \citet{ferguson05}. Finally, the differences on the treatment of overshooting can be identified by comparing the relative descriptions in \citet{bressan12} for the PARSEC models and \citet{scuflaire08} for the CLES models. 

To assess the discrepancies arising from the use of the two different stellar evolutionary models, we analysed a wide sample of {\it CHEOPS} targets with the isochrone placement technique presented in \citet{bonfanti15,bonfanti16} considering both sets of isochrones and tracks. We calculated that differences in age and mass may amount to $\sim$20\% and $\sim$4\%, respectively. Therefore, we considered these values as a reference estimate for the internal precision of the isochrones.

Then, we specifically analysed HD\,108236 to retrieve its mass $M_{\star}$ and age $t_{\star}$. The adopted input parameters were the $T_{\mathrm{eff}}$, [Fe/H], and stellar radius $R_{\mathrm{IRFM},\star}$. Two independent analyses were carried out considering both PARSEC and CLES evolutionary models. The first analysis used the Isochrone placement technique and its interpolating capability applied to pre-computed PARSEC grids of isochrones and tracks to derive the set of stellar masses ($M_{\star,1}\pm\Delta M_{\star,1}$) and ages ($t_{\star,1}\pm\Delta t_{\star,1}$) that best match the input parameters. The second analysis, instead, was performed by directly fitting the input parameters to the CLES stellar models to infer $M_{\star,2}\pm\Delta M_{\star,2}$ and $t_{\star,2}\pm\Delta t_{\star,2}$. To account for model-related uncertainties, we added in quadrature an uncertainty of 20\% in age and of 4\% in mass to the estimates obtained from each set of models. The values obtained from each of the two analyses are listed in Table~\ref{tab:MtPD_LG}. 

From these values we built the corresponding Gaussian probability density functions to then obtain the final estimates of both mass and age. To avoid underestimating the final uncertainties, for each parameter we summed the two Gaussian distributions representing the outputs of the PARSEC and CLES analyses. The median of the combined distribution was assumed as our reference final value, and its corresponding error bars were inferred from the 15.87$^{\mathrm{th}}$ and 84.14$^{\mathrm{th}}$ percentile of the combined distribution, in order to provide the 1$\sigma$ (68.3\%) standard confidence interval. At the end, we obtained $M_{\star}=0.869_{-0.048}^{+0.050}\, M_{\odot}$ and $t_{\star}=6.7_{-5.1}^{+4.0}$ Gyr, as final values for the stellar mass and age, respectively.

\begin{table}
\caption{Stellar mass and age values computed considering the PARSEC and CLES models. Their weighted average consistence has been successfully checked through the $p$-value criterion based on $\chi^2$ tests (see text for details).}
\label{tab:MtPD_LG}
\centering
\begin{tabular}{lllll}
\hline\hline
 \multicolumn{2}{c}{Parameter} & PARSEC & CLES & $p$-value \\
\hline
$M_{\star}$ & [$M_{\odot}$] & $0.853\pm0.043$ & $0.886\pm0.049$ & 0.61 \\
$t_{\star}$ & [Gyr]         & $7.7\pm3.1$     & $4.7\pm5.8$ & 0.62 \\ 
\hline
\end{tabular}
\end{table}

Then we applied a $\chi^2$-test to identify whether results coming from the two different methods (i.e. PARSEC vs CLES) are consistent (null hypothesis), so to check whether their synthesis into single values is indeed a signal of the robustness of the results, rather than a mathematical artefact. To this end, we computed
\begin{equation}
 \bar{\chi}^2 = \displaystyle\sum_{i=1}^2\frac{(x_i-\bar{x})^2}{\sigma_i^2}\,,
 \label{eq:chi2}
\end{equation}
where $x$ denotes the generic variable (either $M_{\star}$ or $t_{\star}$), $\sigma$ its uncertainty, and $\bar{x}$ the median value of the parameter of interest inferred from the combined distribution. We expect $\bar{\chi}^2$ to be drawn from a chi-square distribution $f(\chi_{\nu}^2)$ with $\nu=2-1=1$ degrees of freedom as $\bar{x}$ depends on $x_1$ and $x_2$. Testing the null hypothesis, namely verifying whether the median of the combined distribution properly describes the data, means comparing the $\bar{\chi}^2$ value obtained from Equation~(\ref{eq:chi2}) to a reference $\chi_{\alpha}^2$ value defined by
\begin{equation}
 \int_{\chi_{\alpha}^2}^{+\infty} f(\chi^2)\mathrm{d}\chi^2 = \alpha\,,
 \label{eq:chi2alpha}
\end{equation}
where $\alpha$ is the adopted significance level, which we set equal to 0.05. If $\bar{\chi}^2>\chi_{\alpha}^2$, then we are $100(1-\alpha)\%=95\%$ confident that the null hypothesis is false; otherwise, the null hypothesis is confirmed when
\begin{equation}
 \bar{\chi}^2<\chi_{\alpha}^2 \Leftrightarrow p\mathrm{-value} > \alpha\,,
 \label{eq:pValueH0}
\end{equation}
where
\begin{equation}
 p\mathrm{-value}=\int_{\bar{\chi}^2}^{+\infty} f(\chi^2)\mathrm{d}\chi^2\,.
 \label{eq:pValue}
\end{equation}
These calculations (see $p$-values in Table~\ref{tab:MtPD_LG}) confirm that the two independent derivations of the stellar mass and age are consistent, implying that the median of the combined distribution can be used to assess $M_{\star}$ and $t_{\star}$ from our two sets of measurements. Table~\ref{tab:stellarParam} lists the final adopted stellar parameters.

D20 also derived the stellar fundamental parameters, but employing different approaches: they started either from high-resolution spectroscopy or broad-band photometry to obtain different pairs of ($R_{\star}$, $M_{\star}$). All obtained values agree with our estimates within 1$\sigma$, except for the $M_{\star}$ value that they computed from the mass-radius relation of \citet{torres10}, which is in any case $\sim$2$\sigma$ away also from the other $M_{\star}$ values obtained by D20. By combining the observed spectral energy distribution and the MESA \citep{paxton18} isochrones and stellar tracks \citep[MIST,][]{choi16,dotter16}, D20 find an age of $t_{\star,D20}=5.8\pm4.1$ Gyr, which is consistent with our estimate within $1\sigma$.

\begin{table}
\caption{Properties of HD\,108236 and the methods employed to derive them. See the text for further details.}             
\label{tab:stellarParam}      
\centering                          
\begin{tabular}{lll}        
\hline\hline                 
\multicolumn{3}{c}{HD\,108236} \\    
\hline                        
\multirow{3}{2 cm}{Alternative names} & \multicolumn{2}{l}{TOI-1233} \\ 
 & \multicolumn{2}{l}{HIP 60689} \\ 
 & \multicolumn{2}{l}{Gaia DR2 6125644402384918784} \\
\hline
Parameter & Value & Method \\
\hline
   $V$ [mag] & 9.24 & Simbad \\
   $G$ [mag] & 9.0875 & Simbad \\
   $J$ [mag] & 8.046 & Simbad \\
   $T_{\mathrm{eff}}$ [K] & $5660\pm61$ & spectroscopy \\
   $\log{g}$ [cgs]      & $4.49\pm0.11$ & spectroscopy \\\relax
   [Fe/H] [dex] & $-0.28\pm0.04$ & spectroscopy \\\relax
   [Mg/H] [dex] & $-0.27\pm0.03$ & spectroscopy \\\relax
   [Si/H] [dex] & $-0.29\pm0.02$ & spectroscopy \\
   $d$ [pc] & $64.4\pm0.2$ & Gaia parallax\tablefootmark{(a)}\\
   $\theta$ [mas] & $0.1267\pm0.0012$ & IRFM \\
   $R_{\star}$ [$R_{\odot}$] & $0.877\pm0.008$ & IRFM \\
   $M_{\star}$ [$M_{\odot}$] & $0.869_{-0.048}^{+0.050}$ & isochrones \\
   $t_{\star}$ [Gyr]        & $6.7_{-5.1}^{+4.0}$ & isochrones \\
   $L_{\star}$ [$L_{\odot}$] & $0.708\pm0.047$ & from $R_{\star}$ and $T_{\mathrm{eff}}$\\
   $\rho_{\star}$ [g/cm$^3$] & $1.82\pm0.12$ & from $R_{\star}$ and $M_{\star}$ \\
    
\hline                                   
\end{tabular}
\tablefoot{\tablefoottext{a}{Correction from \citet{stassun18} applied}}
\end{table}

\section{Observations}\label{sec:observations}

\subsection{CHEOPS}\label{sec:observations_cheops}

{\it CHEOPS} \citep{benz20} is an ESA small-class mission, dedicated to observing bright stars ($V\lesssim12$ mag) that are already known to host planets by means of ultra-high-precision photometry. 
The precision of photometric signals is limited by stellar photon noise of 150\,ppm/min for a $V$\,=\,9 magnitude star \citep{broeg13}.
On longer timescales, {\it CHEOPS} achieves a photometric precision of 15.5\,ppm in 6 hours of integration time for a $V$\,$\sim$\,9 mag star \citep{benz20}.

The {\it CHEOPS} instrument is composed of an F/8 Ritchey-Chretien on-axis telescope (30\,cm effective diameter) equipped with a single frame-transfer back-side illuminated charge-coupled device (CCD) detector. The acquired images are defocused to minimise pixel-to-pixel variation effects.

\begin{figure}
    \centering
    \includegraphics[width=\hsize]{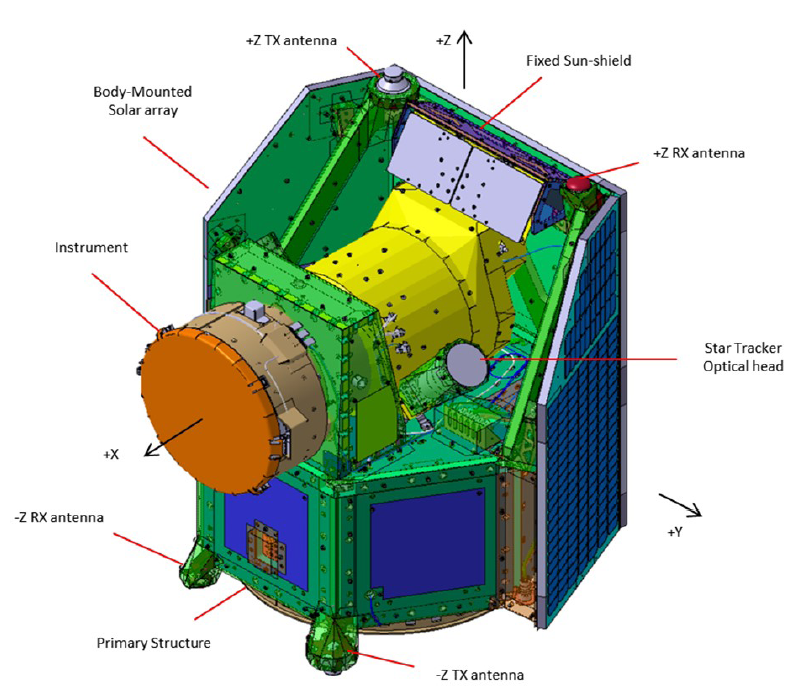}
    \caption{Adopted reference frame of the {\it CHEOPS} spacecraft. The $X$-axis coincides with the line of sight, while the $Z$-axis is antiparallel to the nadir direction. The $x$ and $y$ axes of the CCD coincide with the $-Y$ and $Z$ axes of the spacecraft, respectively. Image taken from \citet{benz20}, courtesy of Airbus Defence and Space, Spain.}
    \label{fig:SCrefFrame}
\end{figure}

The satellite was successfully launched from Kourou (French Guiana) into a $\sim$700\,km altitude solar synchronous orbit on 18 December 2019. The orbit of the spacecraft is nadir-locked (i.e. the $Z$-axis of the spacecraft is antiparallel to the nadir direction, see Figure~\ref{fig:SCrefFrame}) to ensure a thermally stable environment for the payload radiators. During its orbit, the spacecraft rotates around its $X$-axis (the line of sight), and this determines the rotation of the FoV. The angle of rotation around the $X$-axis of the spacecraft is called roll angle, with its zero value occurring when the $Y$-axis of the spacecraft is parallel to the $\mathcal{X}$-$\mathcal{Y}$ plane of the J2000 Earth-centred reference frame. This plane closely approximates the Earth's equatorial plane, coinciding with it on 1 January  2000.
{\it CHEOPS} opened its cover on 29 January 2020 and, after passing the In-Orbit Commissioning (IOC) phase, routine observational operations started on 18 April 2020.

\begin{table*}
\caption{Log of the {\it CHEOPS} observations of HD\,108236. The last column gives the location of the target on the detector.}             
\label{tab:obs_log}      
\centering   
\small
\begin{tabular}{c c c c c c c c}
\hline\hline
Planets & Start date & Duration & Valid points & File key & Efficiency & Exp. time & Location \\ 
 & [UTC] & [h] & [\#] & & [\%] & [s] & $(x; y)$ [px] \\
\hline                    
   c, e & 2020-03-10T18:09:15.9 & 18.33 & 983 & CH\_PR300046\_TG000101\_V0102 & 55 & 42 & $(815; 281)$\\%
   d & 2020-04-28T07:06:11.0 & 18.63 & 770 & CH\_PR100031\_TG015401\_V0102 & 55 & 49 & $(257; 859)$\\%
   b, f & 2020-04-30T17:06:11.0 & 17.04 & 674 & CH\_PR100031\_TG015702\_V0102 & 60 & 49 & $(257; 859)$\\%
\hline                  
\end{tabular}
\end{table*}

\begin{figure}
    \centering
    \includegraphics[width=\hsize]{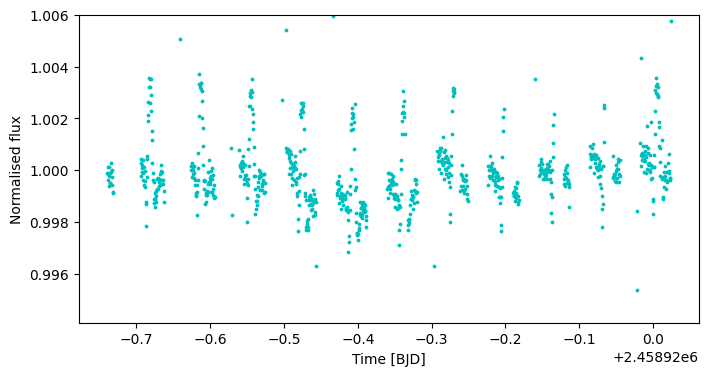} \\
    \includegraphics[width=\hsize]{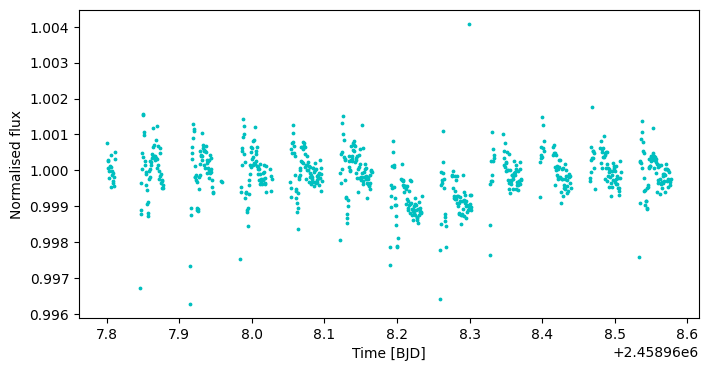} \\
    \includegraphics[width=\hsize]{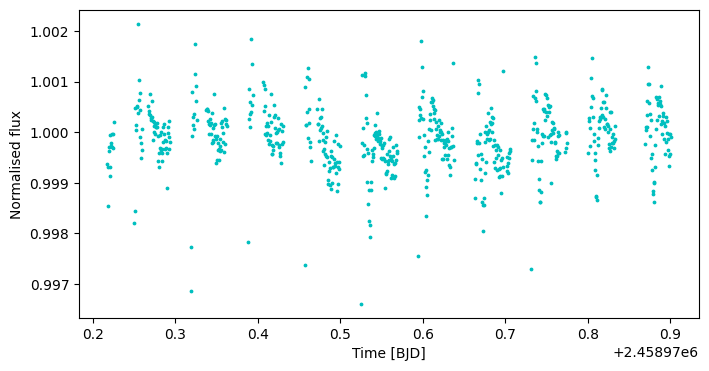}
    \caption{Raw LCs of the three {\it CHEOPS} visits analysed here, as processed by the Data Reduction Pipeline \texttt{v12}. The datasets are presented from top to bottom in chronological order of observation. The periodical light variations, especially visible in the first visit, correlate with the roll angle (see the text for further details).}
    \label{fig:rawLCs}
\end{figure}
Three observation runs, or visits, of HD\,108236 were obtained with {\it CHEOPS} (Figure~\ref{fig:rawLCs}). The first visit, of 18.33\,h duration, was obtained during the IOC phase using exposure times of 42\,s. The other two visits, with a total duration 18.63\,h and 17.04\,h, were obtained during routine operations of the satellite using exposure times of 49\,s. The observations were interrupted by Earth occultations and/or by South Atlantic Anomaly (SAA) crossings, where no data were downlinked, yielding an observing efficiency of 55\%, 55\%, and 60\%, respectively, for each visit. The observing log of the {\it CHEOPS} data is presented in Table~\ref{tab:obs_log}. We also note that the target location on the CCD has changed from March to April. This relocation of the target on the CCD has been done to avoid hot pixels. The project science office constantly monitors the hot pixels status on the detector and if keeping the pre-selected target location implies a significant loss in the expected performances, then the location is changed.

\begin{figure}
    \centering
    \includegraphics[width=\hsize]{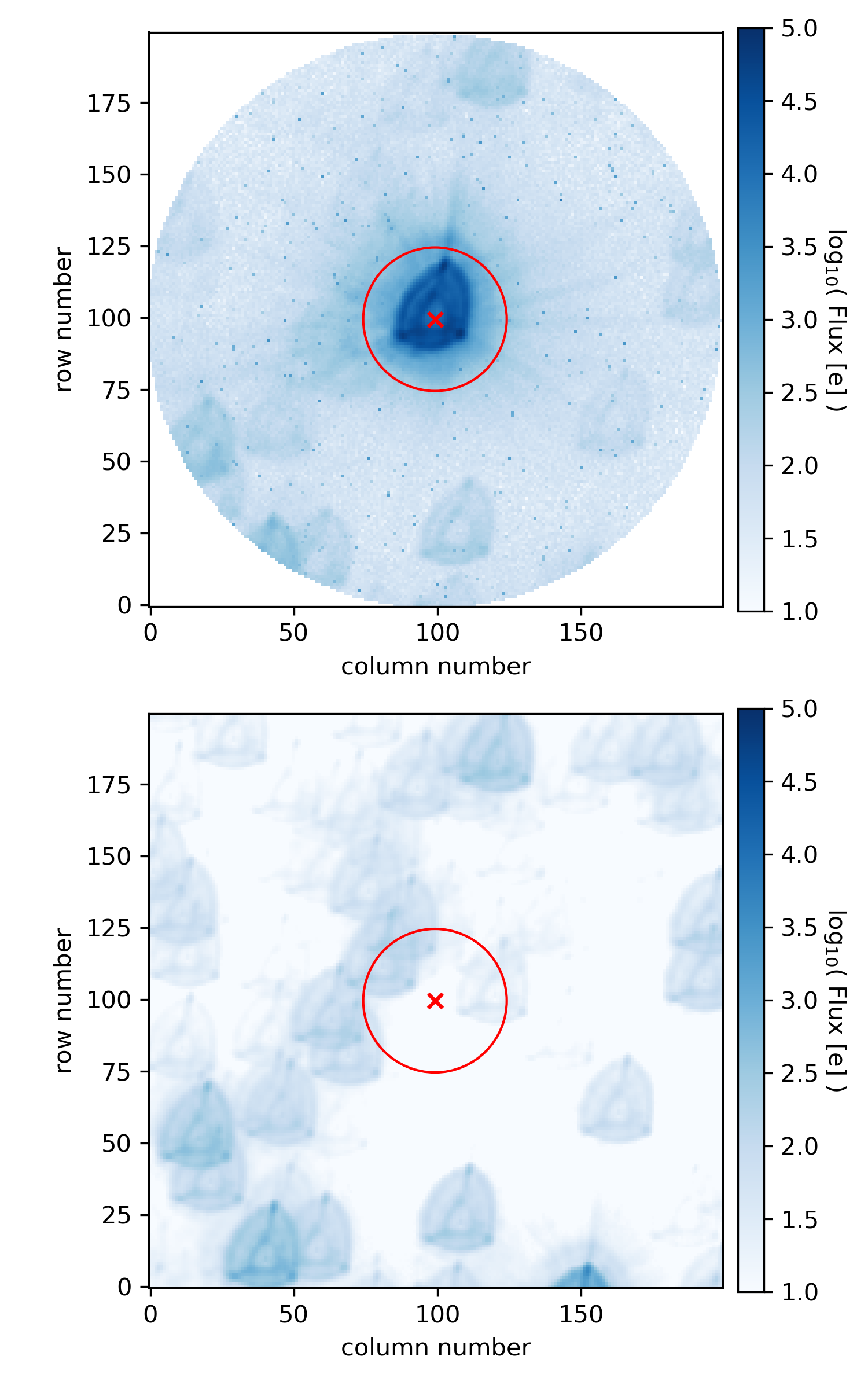}
    \caption{FoV of HD\,108236. Top panel: Real FoV as observed by {\it CHEOPS}. Bottom panel: FoV as inferred from the DRP simulation with the target removed. The red cross indicates the location of the target's PSF, while the photometric aperture is represented by the red circle. The horizontal (column number) and vertical (row number) axes correspond to the $x$ and $y$ axes of the CCD reference frame, respectively. Image scale is 1 arcsec per pixel.}
    \label{fig:fov}
\end{figure}

The raw data were automatically processed by the {\it CHEOPS} Data Reduction Pipeline \citep[DRP \texttt{v12};][]{hoyer2020}. The DRP calibrates and corrects the images for instrumental and environmental effects, and finally performs aperture photometry of the target (Figure~\ref{fig:fov}).
As described in \cite{hoyer2020}, the DRP uses the Gaia catalogue \citep{GaiaCollaboration2018} to simulate the FoV of the observations so to estimate the level of contamination in the photometric aperture. This is achieved by rotating background stars around the spacecraft $X$-axis and/or by the smear trails produced by bright stars in the CCD, in order to mimic the rotating {\it CHEOPS} FoV.
In particular, in \texttt{v12} of the DRP the smear contamination is automatically removed while the background stars' contamination within the aperture (noticeably there is a 4.8\,mag star $\sim$342 arcsec from HD\,108236) is provided as a product to be used for further detrending during the data analysis. Finally, the DRP extracts the photometry using 3 fixed aperture sizes (radii of 22.5, 25 and 30 arcsec) and an extra aperture, the size of which depends on the level of contamination of the FoV. 
In this work we used the LCs obtained with the \texttt{DEFAULT} aperture of 25 arcsec, which results in the smallest root mean square (RMS) in the resulting LCs.

\begin{figure}
\centering
\includegraphics[width=\hsize]{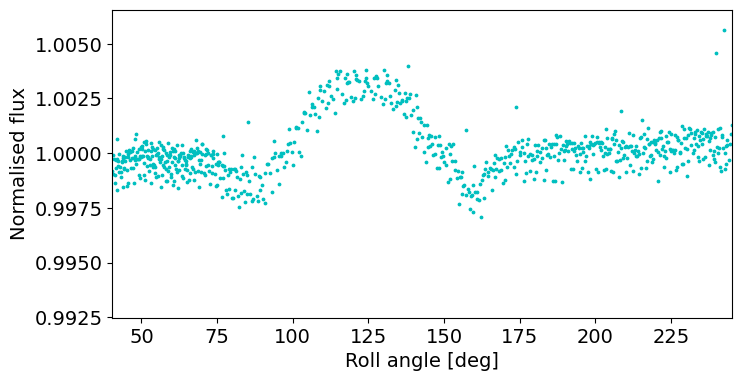} \\
\includegraphics[width=\hsize]{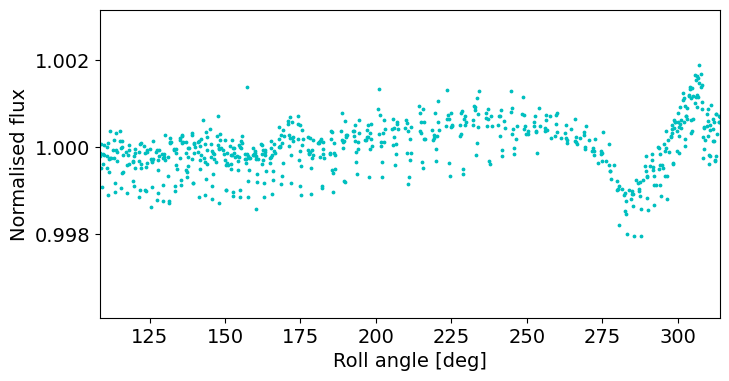} \\
\includegraphics[width=\hsize]{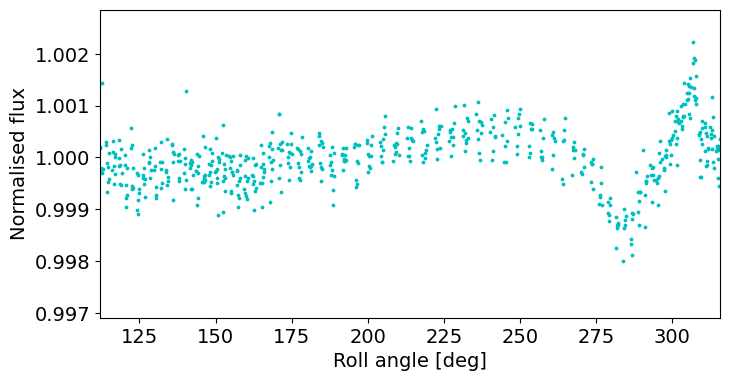}
  \caption{Dependence of flux against roll angle. The three {\it CHEOPS} datasets are presented from top to bottom in chronological order of observation.}
  \label{fig:LCvsRoll}
\end{figure}

\begin{figure}
    \centering
    \includegraphics[width=\hsize]{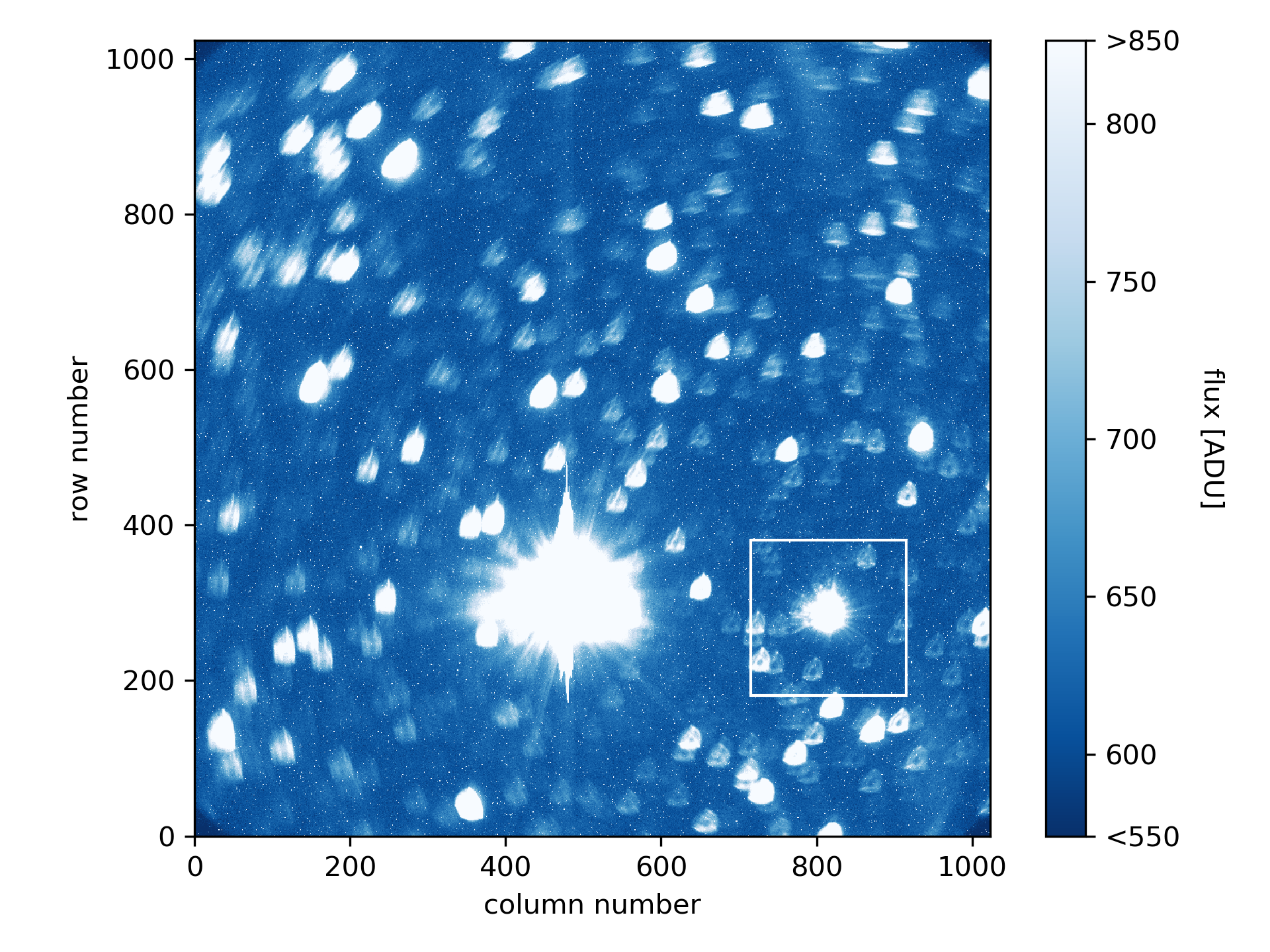}
    \caption{Full frame image of the first \textit{CHEOPS} visit. The white square represents the location and size of the subarray window containing the target. The bright star located on the left of the white square is HD\,108257, which probably produces the internal reflection generating the flux vs roll angle patterns shown in Fig. \ref{fig:LCvsRoll}.}
    \label{fig:frameVisit1}
\end{figure}

\begin{figure*}
    \resizebox{\hsize}{!}{   \includegraphics{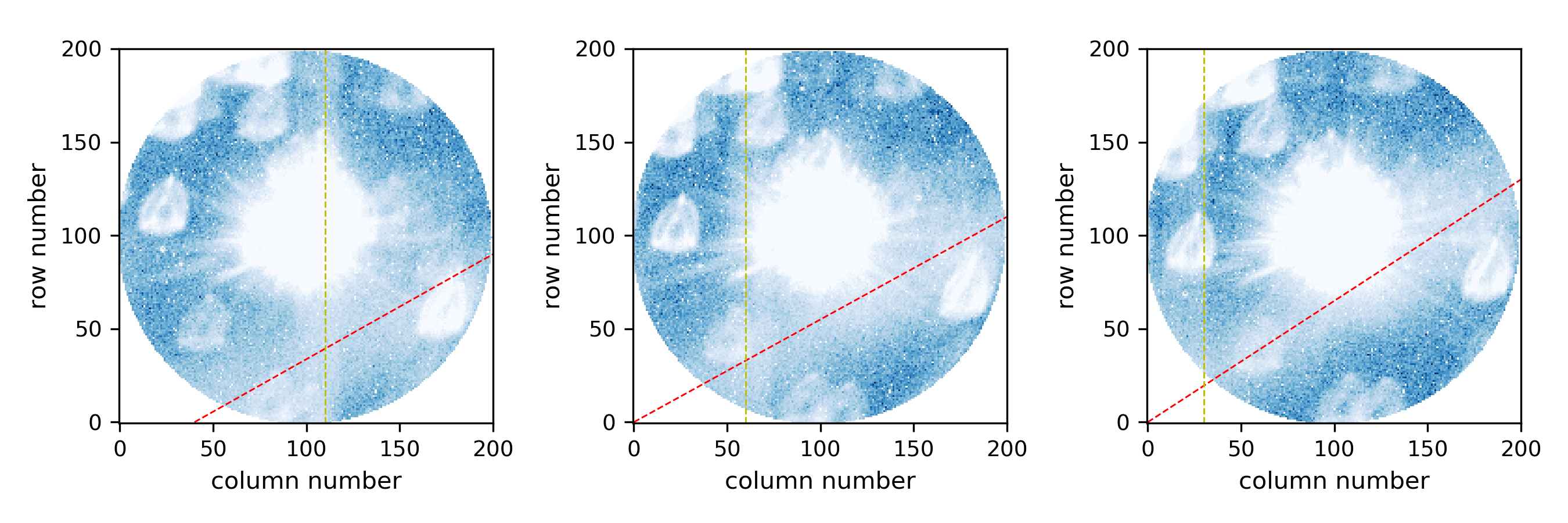}}
    \caption{Set of frames taken from the first {\it CHEOPS} visit. The vertical dashed line indicates the smearing pattern, while the dashed red line indicates another pattern likely produced by the internal reflection of a contaminating source, which could be the nearby bright star HD\,108257.}
    \label{fig:smearBar}
\end{figure*}

We carefully inspected the LCs, looking for possible systematics. It turned out that the stellar flux presents particular patterns against the telescope roll angle in all three datasets, as shown in Figure~\ref{fig:LCvsRoll}. These flux variations against roll angle are likely due to an internal reflection from a very bright nearby star (HD\,108257, $V\,=\,4.8$\,mag), located 342 arcsec away (see Figure~\ref{fig:frameVisit1}). This internal reflection produces a sort of slanted moving bar, which is visible in several frames, as shown in Figure~\ref{fig:smearBar}.

The specific pattern of flux versus roll angle produced by this contamination depends on the location of the target on the CCD. As reported in Table~\ref{tab:obs_log}, the $(x;y)$ coordinates of the target in the first visit differ from those in the following two visits. In addition, the roll angle rotation rate is not constant with time, and it depends on the target's coordinates with respect to the anti-Sun direction. By observing the same target at different epochs, the rotation rate of the spacecraft around the $X$-axis also changes, leading to variations in the rate at which the slanted bar moves. Consistently with this, the flux versus roll angle pattern shown in the top panel of Figure~\ref{fig:LCvsRoll} (mid-March observations) differs from that shown in the middle and bottom panels, which are similar (both observations were taken at the end of April). With this in mind, we detrended the {\it CHEOPS} LCs versus roll angle (see Section \ref{sec:analysis}).

\subsection{TESS}\label{sec:observations_tess}

In addition to the three {\it CHEOPS} datasets, we included the {\it TESS} data analysed by D20. HD\,108236 was observed by {\it TESS} in sectors 10 (26 March 2019 - 22 April 2019) and 11 (22 April 2019 - 21 May 2019). Therefore, combining the {\it TESS} and {\it CHEOPS} data significantly increases the baseline, enabling us to improve the transit ephemerides and the system parameters (see Section \ref{sec:comparative_analysis}). We analysed the {\it TESS} and {\it CHEOPS} data separately, to compare the performances, and together, to better constrain the system parameters.

We used the {\it TESS} data as processed by the Science Processing Operations Center \citep[SPOC;][]{jenkins16} pipeline. In particular, we considered the Pre-search Data Conditioned Simple Aperture Photometry (PDCSAP) flux values with their uncertainties as they are corrected for instrumental variations and represent the best estimate of the intrinsic flux variation of the target.

\section{Data analysis}\label{sec:analysis}

We carried out the LC analysis using \texttt{allesfitter} \citep{guenther20}. Among its features, this framework allows one to model exoplanetary transits using the \texttt{ellc} package \citep{maxted16}, further considering multi-planetary systems, TTVs, and Gaussian processes \citep[GPs;][]{rasmussen05} for the treatment of correlated noise, implemented through the \texttt{celerite} package \citep{foreman17}. The parameters of interest are retrieved considering a Bayesian approach, which may use either the \texttt{emcee} package \citep{foreman13} implementing a MCMC method \citep[see e.g.][]{ford05} to sample the posterior probability distribution, or the Nested Sampling inference algorithm \citep[see e.g.][]{feroz08,feroz19}. In our work, we used the Dynamic Nested Sampling algorithm to have a direct estimate of the Bayesian evidence thanks to the \texttt{dynesty} package \citep{speagle20}. 

Since \texttt{allesfitter} is not able to account for the flux dependence against roll angle seen in the {\it CHEOPS} data, we detrended the flux dependence versus roll angle through a Matérn-3/2 GP before inserting the {\it CHEOPS} LCs into \texttt{allesfitter}. The detrending was done using the \texttt{celerite} package, which gives the GP model and its variance. The GP model was computed using only the out-of-transit data points. Then, new enhanced error bars have been associated with the data points through error propagation accounting for the observational errors and the variance of the GP model.

Throughout the analysis, we assumed Gaussian priors on the following fitted parameters: the mean stellar density $\rho_{\star}=1.82\pm0.12$ g/cm$^3$, derived from our stellar characterisation, and the quadratic limb darkening (LD) coefficients $(q_1, q_2)$, inferred from the \textsc{atlas9} models\footnote{\url{http://kurucz.harvard.edu/grids.html}}. In particular, we derived the $u_1$ and $u_2$ coefficients of the quadratic LD law using the code of \citet{espinoza15}, that performs a cubic spline interpolation within the models according to the same procedure followed by \citet{claret11}. After that, we converted $u_1$ and $u_2$ to the quadratic LD coefficients $q_1$ and $q_2$ required by \texttt{allesfitter} following the relations of \citet{kipping13}, obtaining $(q_1, q_2)=(0.34, 0.27)$ for the {\it TESS} bandpass, and $(q_1, q_2)=(0.46, 0.32)$ for the {\it CHEOPS} bandpass\footnote{The {\it CHEOPS} filter profile (beyond many others, including the {\it TESS} one) may be downloaded as ASCII file e.g. at \url{http://svo2.cab.inta-csic.es/theory/fps/index.php?id=CHEOPS/CHEOPS.band&&mode=browse&gname=CHEOPS&gname2=CHEOPS\#filter}.}. A 1-$\sigma$ uncertainty of 0.05 was attributed to all LD coefficients. We verified that this uncertainty value is conservatively in agreement with the priors estimated by \citet{maxted18}, who discussed the application of the power-2 LD law to the LCs of transiting exoplanets.

For each planet, the fitted parameters were: the ratio of planetary radius over stellar radius $R_p$/$R_{\star}$, the sum of stellar and planetary radius scaled to the orbital semi-major axis $(R_p+R_{\star})$/$a$, the cosine of the orbital inclination $\cos{i_p}$, the transit timing $T_0$, the orbital period $P$, and $\sqrt{e}\cos{\omega}$ and $\sqrt{e}\sin{\omega}$, where $e$ is the orbital eccentricity and $\omega$ is the argument of pericentre. The initial priors used in our fits are listed in Table~\ref{tab:priors}.

\begin{figure}
    \centering
    \includegraphics[width=\hsize]{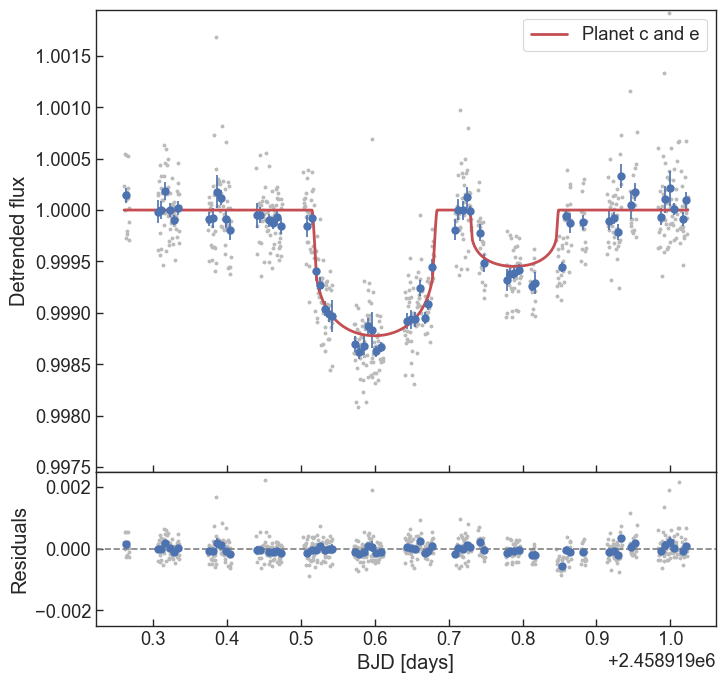}
    \caption{Detrended {\it CHEOPS} LC (first visit, file key CH\_PR300046\_TG000101\_V0102) containing the transits of HD\,108236\,c at $T_{0,\mathrm{c,CH}}\sim$2458919.8\,BJD and HD\,108236\,e at $T_{0,\mathrm{e,CH}}\sim$2458919.6\,BJD. The red line shows the best-fit transit model, while the bottom panel shows the residuals following removal of the transit model.}
    \label{fig:planets_ce}
\end{figure}

\begin{figure}
    \centering
    \includegraphics[width=\hsize]{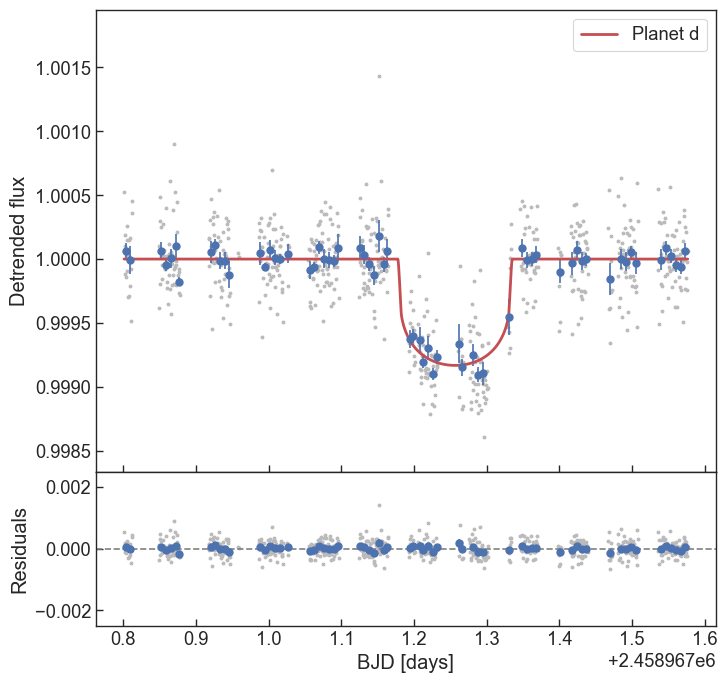}
    \caption{Same as Figure~\ref{fig:planets_ce}, but for HD\,108236\,d (second visit, file key CH\_PR100031\_TG015401\_V0102).}
    \label{fig:planet_d}
\end{figure}

\begin{figure}
    \centering
    \includegraphics[width=\hsize]{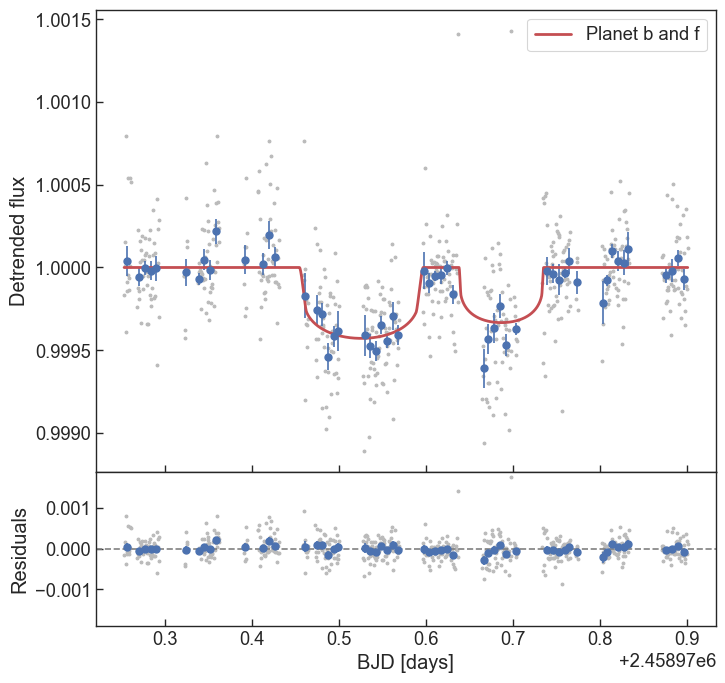}
    \caption{Same as Figure~\ref{fig:planets_ce}, but for HD\,108236\,b and HD\,108236\,f (third visit, file key CH\_PR100031\_TG015702\_V0102). The transit of HD\,108236\,b is that at $T_{0,\mathrm{b,CH}}\sim$2458970.7\,BJD, while the unexpected transit of HD\,108236\,f is that at $T_{0,\mathrm{f,CH}}\sim$2458970.5\,BJD.}
    \label{fig:planets_bf}
\end{figure}

The {\it CHEOPS} LCs, both binned and unbinned, with superimposed best-fit transit models, are shown in Figures~\ref{fig:planets_ce}, \ref{fig:planet_d}, and \ref{fig:planets_bf}. In each Figure presenting {\it TESS} or {\it CHEOPS} LCs, the binned data are shown by combining 12 data points, independently of the exposure times. In the case of the {\it CHEOPS} LCs, each rebinned data point corresponds to 8.4 min in the case of the first visit and 9.8 min in the case of the last two visits. 

During the inspection of the {\it CHEOPS} dataset with file key CH\_PR100031\_TG015702\_V0102 (i.e. last observation), besides finding the expected transit of HD\,108236\,b at $T_{0,\mathrm{b,CH}}\sim$2458970.7\,BJD, we serendipitously detected a transit-like feature (depth $\sim$400\,ppm) occurring $\sim$0.2 days earlier than $T_{0,\mathrm{b,CH}}$ (i.e. $T_{0,\mathrm{f,CH}}\sim$2458970.5\,BJD, see Figure~\ref{fig:planets_bf}). From this moment on, we will refer to this planet as HD\,108236\,f.

Bearing in mind this new detection, we decided to carry out analyses of the available datasets both considering four-planet and five-planet fitting scenarios. To establish whether one model ($\mathcal{M}_1$) is to be preferred over the other ($\mathcal{M}_0$), we computed the Bayes factor $\mathcal{B}_{10}$, which is defined as \citep{kass95}:
\begin{equation}
    \mathcal{B}_{10}=\frac{\mathcal{Z}_1}{\mathcal{Z}_0}\,,
    \label{eq:bayesFactor}
\end{equation}
where $\mathcal{Z}_i$ is the Bayesian evidence (i.e. the marginal likelihood integrated over the entire parameter space) referring to the $i$-th model. The value of $\mathcal{Z}$ is given by the Nested Sampling algorithm, thus $\mathcal{B}_{10}$ could be straightforwardly computed through Equation (\ref{eq:bayesFactor}). The higher the $\mathcal{B}_{10}$, the higher the evidence against $\mathcal{M}_0$ ( i.e. $\mathcal{M}_1$ is to be preferred). Reference values of $\mathcal{B}_{10}$ and corresponding levels of evidence against $\mathcal{M}_0$ are reported in \citet[\S 3.2;][]{kass95}. Here we just recall that very strong evidence against the null hypothesis $\mathcal{M}_0$ (i.e. $\mathcal{M}_1$ is strongly favoured) occurs when $\ln{\mathcal{B}_{10}}>5$.

\section{Results and discussion}\label{sec:results}
\subsection{Additional planets in the system}\label{ssec:additionalPlanets}

\begin{figure}
    \centering
    \includegraphics[width=\hsize]{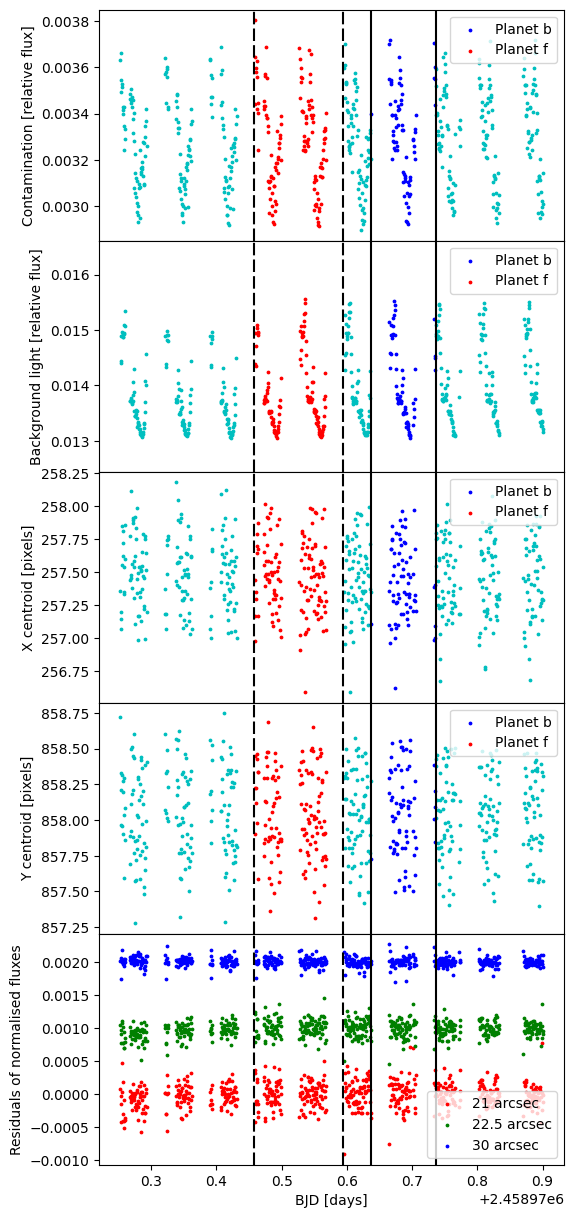}
    \caption{All the represented quantities expressed as a function of time during the third {\it CHEOPS} visit (i.e. the one containing the transits of HD\,108236\,b and HD\,108236\,f). Top two panels: flux of contaminating stars entering the photometric aperture and background light (both relative to the flux of HD\,108236). Third and fourth panels from top: $x$- and $y$-coordinate of the PSF centroid. The transit windows of planets b and f are highlighted in blue and red, respectively, while the out-of-transit points are in cyan. Bottom panel: residuals of the normalised fluxes (shifted vertically for visualisation purposes) obtained after subtracting the reference normalised flux computed considering the \texttt{DEFAULT} aperture to the normalised fluxes computed considering the other three available apertures. No correlation is present between flux and aperture size. Transit windows of planets b and f are marked by solid and dashed vertical lines, respectively.}
    \label{fig:fluxCorr}
\end{figure}

The detected transit-like signal visible in Figure~\ref{fig:planets_bf} at $T_{0,\mathrm{f,CH}}\sim$2458970.5\,BJD does not show any correlation with the flux of field stars that contaminate the aperture photometry \citep[estimated as explained in][]{hoyer2020}, nor with background light, nor with aperture size. In addition, the temporal stability of the $(x; y)$ coordinates of the PSF centroid suggests the absence of any PSF jumps, hence no new hot pixels have appeared inside the PSF area during the observation (see Figure~\ref{fig:fluxCorr}). Furthermore, by analysing the raw data, the DRP team confirmed that this feature cannot be ascribed to telegraphic pixels (i.e. pixels that occasionally blink and twinkle), nor to cosmic rays or any spacecraft instrumental metrics.
 
\begin{figure}
    \centering
    \includegraphics[width=\hsize]{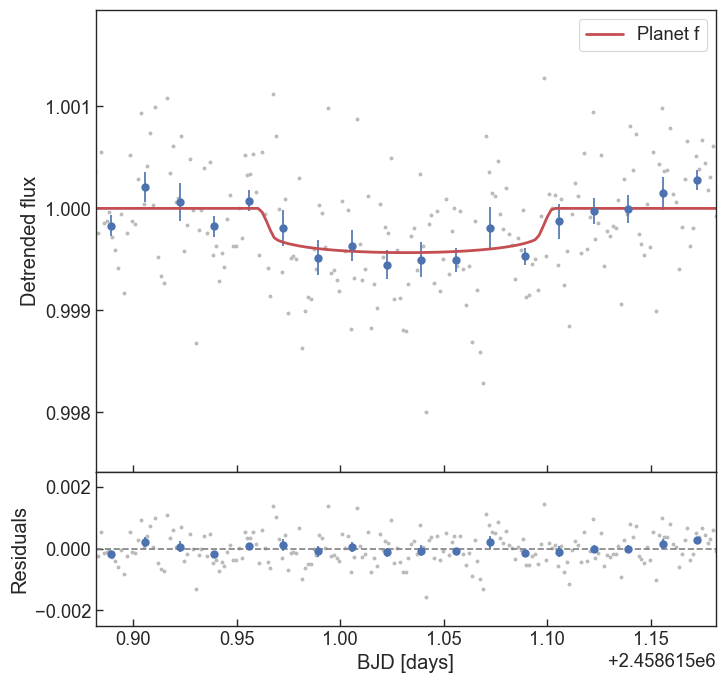}
    \caption{Portion of the {\it TESS} sector 11 LC. Data are binned on a timescale of 24 min. The red solid line shows the transit model for HD\,108236\,f. Neglecting the presence of HD\,108236\,f and performing a four-planet fit, the RMS of the binned residuals would be 154 ppm. Instead, a five-planet fit reduces the RMS of the binned residuals to 130 ppm.}
    \label{fig:TESS2planet_f}
\end{figure}

Therefore, given the transit-like nature of the signal, we looked for similar features, in terms of both transit depth and duration, in the available {\it TESS} LCs. Indeed, we spotted a similar signal in the second {\it TESS} LC (sector 11) at $T_{0\mathrm{,f,TE11}}$\,$\sim$\,2458616.04 BJD by visual inspection (Figure~\ref{fig:TESS2planet_f}). We applied both a four-planet and five-planet fit to these {\it TESS} data and compared the RMS of the residuals binned on a 24 min timescale within the [2458615.9, 2458616.2] BJD window, which contains the supposed transit. We obtained RMS values of 130 ppm and 154 ppm for the five-planet and four-planet scenarios, respectively. The lowest RMS value in the five-planet case shows how the transit model may justify the flux variability; as a term of comparison the RMS of the binned residuals over the entire sector 11 is 152 ppm.

\begin{figure}
    \centering
    \includegraphics[width=\hsize]{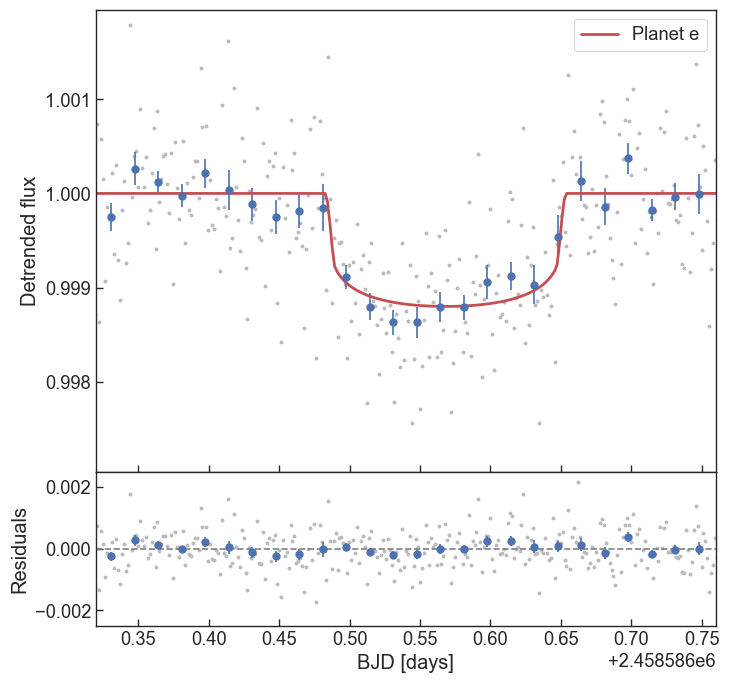}
    \caption{Portion of the {\it TESS} sector 10 LC. The solid red line represents the model of the transit of HD\,108236\,e considering the four-planets scenario: a residual variability in the data points remains unexplained. Data are binned on a timescale of 24 min; the RMS of the binned residuals over the transit window is 168 ppm.}
    \label{fig:TESS1planet_efit4}
\end{figure}

\begin{figure}
    \centering
    \includegraphics[height=0.905\textheight]{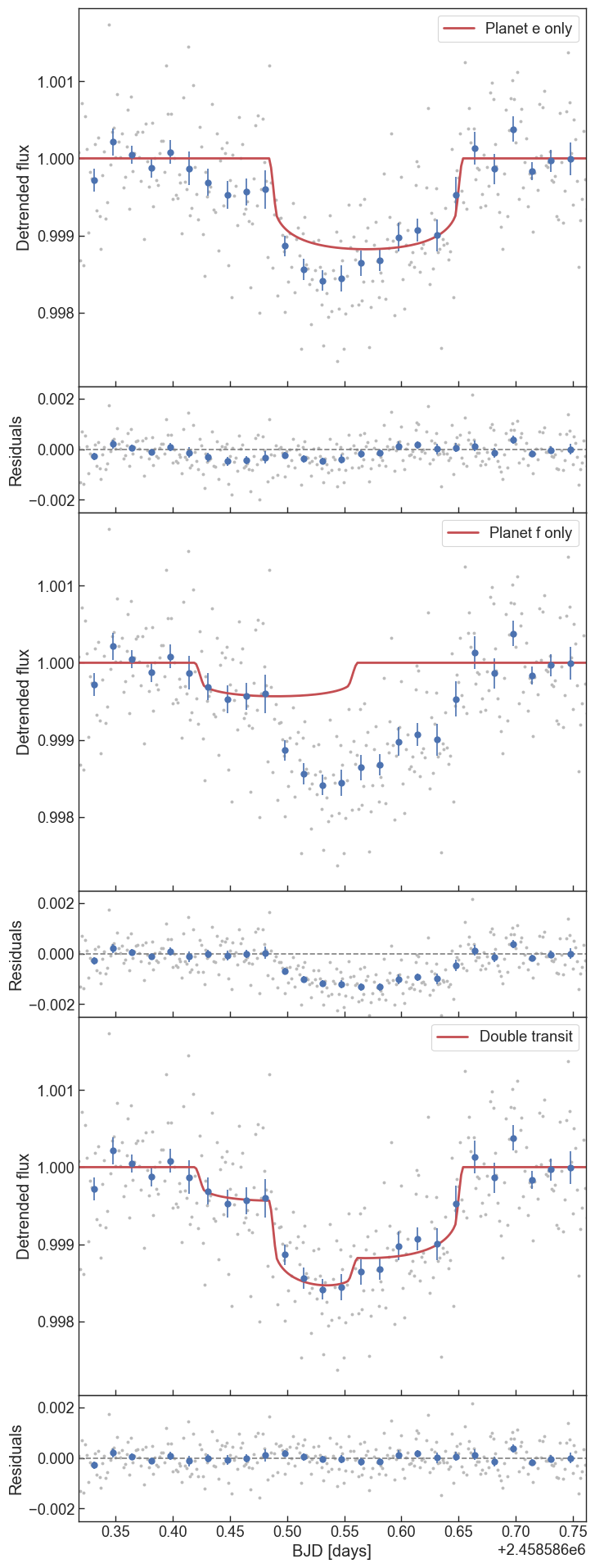}
    \caption{Same as Figure~\ref{fig:TESS1planet_efit4}, but considering the five-planet fit. The top and middle panels show the transits of HD\,108236\,e and HD\,108236\,f separately, while the bottom panel shows the results obtained with the combined model. The RMS of the binned residuals over the transit window is 137 ppm, which is lower than 168 ppm retrieved in the four-planet scenario, where HD\,108236\,f is absent.}
    \label{fig:TESS1double_ef}
\end{figure}

Then, we carefully inspected the {\it TESS} sector 10 LC as well, looking for further undetected signals compatible with a d$F$\,$\sim$\,400 ppm. We noticed that $\sim$\,29.5 days before the supposed transit of HD\,108236\,f, HD\,108236\,e is transiting, but the assumption of four planets keeps some additional flux variability when the transit model is subtracted from the data points (see Figure~\ref{fig:TESS1planet_efit4}). We wondered whether the structures seen in the signal could be ascribed to a further undetected transit, that is whether the layout of the data points could be justified by a double transit.
By including the transit model of HD\,108236\,f, we significantly improve the quality of the fit as shown in Figure~\ref{fig:TESS1double_ef}. Considering the temporal window [2458586.4, 2458586.7] BJD, the RMS of the binned residuals on a 24 minutes timescale is 137 ppm for the five-planet scenario versus 168 ppm for the four-planet scenario. As a term of comparison, the RMS of the binned residuals over the entire sector 10 LC is 158 ppm. The presence of HD\,108236\,f also affects the transit depth of HD\,108236\,e, d$F_{\mathrm{e}}$. From the five-planet scenario we inferred d$F_{\mathrm{e}}=1038_{-28}^{+30}$ ppm, while the four-planet scenario yields to a deeper transit with d$F_{\mathrm{e}}=1056_{-36}^{+37}$ ppm, as expected.

The temporal difference between these two transits of HD\,108236\,f in the two {\it TESS} sectors gives a candidate value for the orbital period, which also agrees with the transit epoch of the signal detected in the {\it CHEOPS} visit. Therefore, we propose $P=29.54$ days as a possible candidate value. 

D20 investigated whether additional planets may be present in the system and proposed a possible candidate with a period of 10.9113 days, with $T_0=2458570.6781$\,BJD and a transit depth of 230 ppm. However, they cautioned that this is only a candidate because the false alarm probability of 0.01 is influenced by the detrending method and the detected features might be instrumental in origin. As the {\it TESS} dataset cannot give a definitive answer, we looked for possible signals of this candidate in the {\it CHEOPS} LCs. Propagating its ephemeris, we would expect a transit of this candidate planet at $T_0=2458919.839$\,BJD, hence blended with the transit of HD\,108236\,c. We ran a further analysis of the {\it CHEOPS} LC, covering the transit of HD\,108236\,c considering a scenario with six planets and compared the results with the favoured scenario with five planets (null hypothesis), obtaining $\ln{\mathcal{B}_{65}}=-10.6\Rightarrow \ln{\mathcal{B}_{56}}=10.6>5$, which means that there is a strong evidence for not rejecting the null hypothesis. On the one hand, the Bayes factor strongly disfavours the presence of a planet at $\sim$10.9 days. On the other hand, the putative transit would be shallow and blended with the transit of HD\,108236\,c. Therefore, further LCs are needed to give a definitive answer.

\subsection{Comparative photometric analysis}\label{sec:comparative_analysis}

We carried out the analysis considering the {\it TESS} and {\it CHEOPS} datasets separately ('TESS-only' and 'CHEOPS-only' approaches) and combined ('TESS+CHEOPS' approach). As the CHEOPS-only approach involved the analysis of one single transit per planet, we imposed a normal prior on the orbital periods based on the results obtained from the TESS-only approach. 
For each approach, we fitted the data considering both four and five planets. The four-planet fit considered the four planets from HD\,108236\,b to HD\,108236\,e already detected by D20, while the five-planet fit also included the planet HD\,108236\,f with an orbital period of $\sim$\,29.5\,days. Here we recall that, according to our discussion in Section~\ref{ssec:additionalPlanets}, there should be just two transits of HD\,108236\,f in the whole {\it TESS} dataset, and one is blended with a transit of HD\,108236\,e.

The Bayesian factors $\mathcal{B}_{54}$ obtained from the different approaches, where we tested the five-planet scenario against the four-planet scenario, are reported in Table~\ref{tab:BayesFactors}.
The $\ln{\mathcal{B}_{54}}$ Bayesian indicator (which is always greater than 5) strongly favours the scenario with five transiting planets.

\begin{table}
\caption{Natural logarithm of the Bayes factors obtained when testing the five-planet scenario against the four-planet scenario for each of our three approaches.}             
\label{tab:BayesFactors}      
\centering                          
\begin{tabular}{l c c c}        
\hline\hline                 
 & TESS-only & CHEOPS-only & TESS+CHEOPS \\    
\hline                        
   $\ln{\mathcal{B}_{54}}$ & 10.8 & 13.4 & 13.1 \\      
\hline                                   
\end{tabular}
\end{table}

Supported by the Bayesian evidence, we consider the five-planet scenario to be the most appropriate description of the data. Table~\ref{tab:parameters} summarises the results obtained for this scenario. At first, we performed all analyses considering the orbital eccentricity $e$ as a free parameter, as also done by D20. Then, we analysed the dynamical stability of our TESS+CHEOPS solution (see Section~\ref{ssec:dynamical_analysis} for further details), finding that the lack of a constraint on $e$ leads to a severely unstable system. Therefore, we repeated the TESS+CHEOPS analysis setting $e\lesssim0.1$, which is also supported by the dynamical stability analysis of D20 (see their Figure 16). The last column of Table~\ref{tab:parameters} presents our reference results, accounting for the constraint on $e$ and considering all available datasets. 
We emphasise that in Table~\ref{tab:parameters} all reported transit depths are intended to be $\mathrm{d}F\equiv \left(\frac{R_p}{R_{\star}}\right)^2$ to make them comparable between {\it CHEOPS} and {\it TESS}, that is to say independent of the instrumental bandpass and thus specific limb darkening.

\begin{table*}
\caption{HD\,108236 system parameters obtained from analyses of the {\it TESS} and {\it CHEOPS} LCs, in comparison to what was found by \citet{Daylan2020}. The columns labelled as TESS-only, CHEOPS-only, and TESS+CHEOPS refer to the results obtained considering five transiting planets in the system. All results have been obtained considering $e$ as a free parameter, except for the last column for which $e$ has been set to be smaller than 0.1. The TESS+CHEOPS analysis with $e<0.1$ gives our final adopted solution for the system.}             
\label{tab:parameters}      
\centering          
\footnotesize
\begin{tabular}{ll ll ll ll}
\hline\hline       
\multicolumn{3}{c}{Parameters} & \citet{Daylan2020} & TESS-only & CHEOPS-only & \multicolumn{2}{c}{TESS+CHEOPS} \\
\multicolumn{3}{c}{} & $e$ free & $e$ free & $e$ free & $e$ free & $e< 0.1$ \\
\hline
\multicolumn{8}{l}{Planet b: 12 {\it TESS} transits; 1 {\it CHEOPS} transit} \\
\hline
   Eclipse timing\tablefootmark{(a)} & $T_0$ & [BJD] & $8572.1128_{-0.0036}^{+0.0031}$ & $8598.6796_{-0.0025}^{+0.0019}$ & $8947.9067_{-0.0020}^{+0.0055}$ & 
   $8773.2959_{-0.0017}^{+0.0012}$ & $8773.2964\pm0.0013$ \\  
   Period         & $P$ & [days]  & $3.79523_{-0.00044}^{+0.00047}$ & $3.79517_{-0.00037}^{+0.00040}$ & $3.79527_{-0.00013}^{+0.00013}$ & 
   $3.795954_{-0.000034}^{+0.000029}$ & $3.795963_{-0.000034}^{+0.000025}$ \\
   Transit depth\tablefootmark{(b)} & d$F$ & [ppm] & $268\pm31$ & $267_{-24}^{+26}$ & $337_{-45}^{+36}$ &
   $273\pm17$ & $285\pm17$ \\
   Transit duration & $W$ &[h]& $2.30_{-0.11}^{+0.16}$ & $2.35_{-0.28}^{+0.17}$ & $2.09_{-0.29}^{+0.35}$ &
   $2.46_{-0.15}^{+0.12}$ & $2.384\pm0.093$ \\
   Impact parameter & $b$ & [$R_{\star}$] & $0.38\pm0.24$ & $0.48_{-0.15}^{+0.17}$ & $0.65_{-0.19}^{+0.11}$ &
   $0.43\pm0.12$ & $0.452_{-0.096}^{+0.071}$ \\
   Semi-major axis & $a$ & [AU] & $0.0469\pm0.0017$ & $0.04505\pm0.00077$ & $0.04454_{-0.0012}^{+0.00083}$ &
   $0.04436\pm0.00081$ & $0.04527_{-0.00088}^{+0.00096}$ \\
   Orbital inclination & $i_p$ & [$^{\circ}$] & $87.88_{-0.87}^{+1.30}$ & $87.47_{-0.59}^{+0.65}$ & $86.04_{-0.44}^{+0.50}$ &
   $87.51_{-0.50}^{+0.58}$ & $87.59_{-0.35}^{+0.47}$ \\
   Eccentricity & $e$ &  & $0.20_{-0.14}^{+0.30}$ & $0.130_{-0.094}^{+0.18}$ & $0.22_{-0.17}^{+0.19}$ &
   $0.20_{-0.11}^{+0.15}$ & $0.045_{-0.033}^{+0.039}$ \\
   Arg. of pericentre & $\omega$ & [$^{\circ}$] & $190\pm140$ & $262_{-230}^{+77}$ & $49_{-32}^{+130}$ &
   $166_{-30.}^{+21}$ & $126_{-62}^{+51}$ \\
   Radius & $R_p$ & [$R_{\oplus}$]    & $1.586\pm0.098$ & $1.563_{-0.072}^{+0.078}$ & $1.756_{-0.12}^{+0.096}$ &
   $1.579\pm0.052$ & $1.615\pm0.051$ \\
\hline
\multicolumn{8}{l}{Planet c: 8 {\it TESS} transits; 1 {\it CHEOPS} transit} \\
\hline
   Eclipse timing\tablefootmark{(a)} & $T_0$ & [BJD] & $8572.3949_{-0.0020}^{+0.0025}$ & $8597.2096_{-0.0015}^{+0.0015}$ & $8944.6086_{-0.0021}^{+0.0022}$ &
   $8770.90623_{-0.00071}^{+0.00071}$ & $8770.90586_{-0.00078}^{+0.00078}$ \\  
   Period         & $P$ & [days]  & $6.20370_{-0.00052}^{+0.00064}$ & $6.20381_{-0.00043}^{+0.00037}$ & $6.20380_{-0.00012}^{+0.00013}$ &
   $6.203468_{-0.000049}^{+0.000035}$ & $6.203449_{-0.000046}^{+0.000049}$ \\
   Transit depth\tablefootmark{(b)} & d$F$ & [ppm] & $455_{-35}^{+40}$ & $429_{-23}^{+27}$ & $524_{-25}^{+32}$ &
   $483_{-20}^{+21}$ & $469_{-22}^{+21}$ \\
   Transit duration & $W$ & [h] & $2.913\pm0.095$ & $2.948_{-0.13}^{+0.099}$ & $2.81_{-0.19}^{+0.14}$ &
   $2.915\pm0.10$ & $2.86_{-0.11}^{+0.10}$ \\
   Impact parameter & $b$ & [$R_{\star}$] & $0.33_{-0.21}^{+0.25}$ & $0.30\pm0.12$ & $0.43_{-0.13}^{+0.12}$ &
   $0.441_{-0.079}^{+0.063}$ & $0.455_{-0.083}^{+0.069}$ \\
   Semi-major axis & $a$ & [AU] & $0.0651\pm0.0024$ & $0.0638_{-0.0011}^{+0.0011}$ & $0.06372_{-0.00082}^{+0.00086}$ &
   $0.0612\pm0.0010$ & $0.0620\pm0.0012$ \\
   Orbital inclination & $i_p$& [$^{\circ}$] & $88.72_{-0.74}^{+0.82}$ & $88.87\pm0.42$ & $88.29_{-0.41}^{+0.47}$ &
   $88.28_{-0.22}^{+0.27}$ & $88.30_{-0.23}^{+0.29}$ \\
   Eccentricity & $e$ &  & $0.18_{-0.14}^{+0.34}$ & $0.069_{-0.050}^{+0.099}$ & $0.098_{-0.063}^{+0.081}$ &
   $0.054_{-0.038}^{+0.073}$ & $0.034_{-0.025}^{+0.042}$ \\
   Arg. of pericentre & $\omega$ & [$^{\circ}$] & $210\pm120$ & $164_{-100.}^{+120}$ & $89\pm53$ &
   $97_{-75}^{+230}$ & $211_{-110}^{+44}$ \\
   Radius & $R_p$ & [$R_{\oplus}$]    & $2.068_{-0.091}^{+0.100}$ & $1.981_{-0.058}^{+0.064}$ & $2.190_{-0.056}^{+0.071}$ &
   $2.102\pm0.050$ & $2.071\pm0.052$ \\
\hline
\multicolumn{8}{l}{Planet d: 4 {\it TESS} transits; 1 {\it CHEOPS} transit} \\
\hline
   Eclipse timing\tablefootmark{(a)} & $T_0$ & [BJD] & $8571.3368_{-0.0013}^{+0.0015}$ & $8599.68768_{-0.00095}^{+0.0013}$ & $8954.0698_{-0.0030}^{+0.0096}$ &
   $8769.79772_{-0.00083}^{+0.0014}$ & $8769.79682_{-0.00090}^{+0.00090}$ \\  
   Period         & $P$ & [days]  & $14.17555_{-0.00110}^{+0.00099}$ & $14.1756\pm0.0011$ & $14.17549_{-0.00048}^{+0.00053}$ &
   $14.175748_{-0.000067}^{+0.00010}$ & $14.175685_{-0.00010}^{+0.000083}$ \\
   Transit depth\tablefootmark{(b)} & d$F$ & [ppm]   & $787\pm53$ & $740_{-33}^{+38}$ & $788_{-58}^{+59}$
   & $754\pm29$ & $705_{-34}^{+31}$ \\
   Transit duration & $W$ & [h] & $3.734_{-0.049}^{+0.066}$ & $4.022_{-0.26}^{+0.088}$ & $2.71_{-0.32}^{+1.3}$ &
   $3.954_{-0.11}^{+0.097}$ & $3.85_{-0.13}^{+0.18}$ \\
   Impact parameter & $b$ & [$R_{\star}$] & $0.35_{-0.21}^{+0.19}$ & $0.25_{-0.14}^{+0.19}$ & $0.776_{-0.61}^{+0.067}$ &
   $0.327\pm0.075$ & $0.423_{-0.13}^{+0.070}$ \\
   Semi-major axis & $a$ & [AU] & $0.1131\pm0.0040$ & $0.1093_{-0.0022}^{+0.0019}$ & $0.1100_{-0.0018}^{+0.0021}$ &
   $0.1087\pm0.0020$ & $0.1074\pm0.0023$ \\
   Orbital inclination & $i_p$ & [$^{\circ}$] & $89.22_{-0.38}^{+0.45}$ & $89.40_{-0.37}^{+0.33}$ & $88.462_{-0.081}^{+0.43}$ &
   $89.25_{-0.15}^{+0.16}$ & $89.06_{-0.14}^{+0.27}$ \\
   Eccentricity & $e$ &  & $0.17_{-0.12}^{+0.30}$ & $0.17_{-0.11}^{+0.18}$ & $0.24_{-0.17}^{+0.51}$ &
   $0.071_{-0.036}^{+0.038}$ & $0.055_{-0.035}^{+0.038}$ \\
   Arg. of pericentre & $\omega$ & [$^{\circ}$] & $190_{-130}^{+140}$ & $167_{-70.}^{+24}$ & $224_{-67}^{+61}$ &
   $79_{-41}^{+46}$ & $151_{-33}^{+46}$ \\
   Radius & $R_p$ & [$R_{\oplus}$] & $2.72\pm0.11$ & $2.603_{-0.065}^{+0.072}$ & $2.684_{-0.099}^{+0.10}$ &
   $2.626\pm0.057$ & $2.539_{-0.065}^{+0.062}$ \\
\hline
\multicolumn{8}{l}{Planet e: 2 {\it TESS} transits; 1 {\it CHEOPS} transit} \\
\hline
   Eclipse timing\tablefootmark{(a)} & $T_0$ & [BJD] & $8586.5677_{-0.0014}^{+0.0014}$ & $8606.1591_{-0.0010}^{+0.0014}$ & $8939.1926_{-0.0018}^{+0.0037}$ &
   $8782.46937_{-0.00062}^{+0.00067}$ & $8782.46935_{-0.00069}^{+0.00078}$ \\  
   Period         & $P$ & [days]  & $19.5917^{+0.0022}_{-0.0020}$ & $19.5899\pm0.0018$ & $19.59087_{-0.00047}^{+0.00053}$ &
   $19.589970_{-0.000074}^{+0.000074}$ & $19.590025_{-0.000089}^{+0.000083}$ \\
   Transit depth\tablefootmark{(b)} & d$F$ & [ppm]   & $1043^{+78}_{-71}$ & $1003_{-50}^{+55}$ & $1049_{-47}^{+39}$ &
   $1036_{-25}^{+31}$ & $1038_{-28}^{+30}$ \\
   Transit duration & $W$ & [h] & $4.013^{+0.080}_{-0.057}$ & $4.12_{-0.30}^{+0.26}$ & $4.22_{-0.18}^{+0.20}$ &
   $4.38_{-0.19}^{+0.15}$ & $4.20_{-0.13}^{+0.13}$ \\
   Impact parameter & $b$ & [$R_{\star}$] & $0.36^{+0.20}_{-0.23}$ & $0.47_{-0.15}^{+0.12}$ & $0.464_{-0.11}^{+0.080}$ &
   $0.374\pm0.10$ & $0.426_{-0.077}^{+0.059}$ \\
   Semi-major axis & $a$ & [AU] & $0.1400\pm0.0052$ & $0.1362\pm0.0026$ & $0.1333\pm0.0017$ &
   $0.1341\pm0.0020$ & $0.1367_{-0.0020}^{+0.0022}$ \\
   Orbital inclination & $i_p$ & [$^{\circ}$] & $89.32^{+0.42}_{-0.30}$ & $89.09_{-0.20}^{+0.19}$ & $89.14_{-0.11}^{+0.17}$ &
   $89.28_{-0.14}^{+0.18}$ & $89.245_{-0.086}^{+0.12}$ \\
   Eccentricity & $e$ &  & $0.20^{+0.30}_{-0.13}$ & $0.29_{-0.20}^{+0.41}$ & $0.20_{-0.15}^{+0.27}$ &
   $0.119_{-0.060}^{+0.086}$ & $0.058_{-0.036}^{+0.032}$ \\
   Arg. of pericentre & $\omega$ & [$^{\circ}$] & $170_{-130}^{+150}$ & $194_{-44}^{+16}$ & $192_{-21}^{+13}$ &
   $136_{-42}^{+28}$ & $128_{-54}^{+37}$ \\
   Radius & $R_p$ & [$R_{\oplus}$] & $3.12^{+0.13}_{-0.12}$ & $3.029_{-0.081}^{+0.088}$ & $3.096_{-0.070}^{+0.066}$ &
   $3.080_{-0.048}^{+0.054}$ & $3.083\pm0.052$ \\
\hline
\multicolumn{8}{l}{Planet f: 2 {\it TESS} transits; 1 {\it CHEOPS} transit} \\
\hline
   Eclipse timing\tablefootmark{(a)} & $T_0$ & [BJD] &  & $8616.0418_{-0.0078}^{+0.0072}$ & $8970.5257_{-0.0038}^{+0.0023}$ &
   $8793.2785_{-0.0022}^{+0.0028}$ & $8793.2786_{-0.0019}^{+0.0021}$ \\  
   Period         & $P$ & [days] & & $29.5461_{-0.0096}^{+0.0089}$ & $29.54255_{-0.00069}^{+0.00069}$ &
   $29.54075_{-0.00045}^{+0.00037}$ & $29.54115_{-0.00042}^{+0.00033}$ \\
   Transit depth\tablefootmark{(b)} & d$F$ & [ppm] & & $366_{-59}^{+61}$ & $416_{-18}^{+21}$ &
   $474_{-32}^{+28}$ & $445_{-24}^{+21}$ \\
   Transit duration & $W$ & [h] & & $4.90_{-1.1}^{+0.33}$ & $4.05_{-0.33}^{+0.35}$ &
   $3.39_{-0.28}^{+0.36}$ & $3.27_{-0.15}^{+0.13}$ \\
   Impact parameter & $b$ & [$R_{\star}$] & & $0.38_{-0.22}^{+0.33}$ & $0.665_{-0.082}^{+0.067}$ &
   $0.793_{-0.061}^{+0.041}$ & $0.808\pm0.022$ \\
   Semi-major axis & $a$ & [AU] & & $0.1773\pm0.0036$ & $0.1759\pm0.0037$ &
   $0.1750\pm0.0041$ & $0.1758_{-0.0038}^{+0.0041}$ \\
   Orbital inclination & $i_p$ & [$^{\circ}$] & & $89.27_{-0.26}^{+0.40}$ & $88.873_{-0.096}^{+0.066}$ &
   $88.910_{-0.078}^{+0.074}$ & $88.963\pm0.042$ \\
   Eccentricity & $e$ &  & & $0.38\pm0.23$ & $0.245\pm0.092$ &
   $0.17_{-0.12}^{+0.19}$ & $0.051_{-0.034}^{+0.036}$ \\
   Arg. of pericentre & $\omega$ & [$^{\circ}$] & & $167_{-67}^{+38}$ & $91\pm29$ &
   $230_{-210}^{+120}$ & $275_{-63}^{+45}$ \\
   Radius & $R_p$ & [$R_{\oplus}$] & & $1.83\pm0.16$ & $1.951_{-0.049}^{+0.054}$ &
   $2.082_{-0.073}^{+0.065}$ & $2.017_{-0.057}^{+0.052}$ \\
\hline
\end{tabular}
\tablefoot{
\tablefoottext{a}{Epoch shifted by $-$2\,450\,000. Time standard is Terrestrial Time (TT).}\quad
\tablefoottext{b}{d$F\equiv\left(\frac{R_p}{R_{\star}}\right)^2$}
}
\end{table*}

As a sanity check, we also carried out an independent joint analysis of the {\it TESS} and {\it CHEOPS} LCs with the code \texttt{pyaneti} \citep{Barragan2019}, which estimates the transit parameters using a Bayesian approach. We imposed uniform priors for all fitted parameters. We sampled for the mean stellar density $\rho_\star$ and recovered the scaled semi-major axis for each planet ($a/R_\star$) using Kepler’s third law \citep{Winn2010}. The joint modelling of the transit LCs of the five planets provides a mean stellar density of $\rho_\star$\,=\,2.11$^{+0.18}_{-0.20}$\,g\,cm$^{-3}$, which agrees with the density of 1.82\,$\pm$\,0.11\,g\,cm$^{-3}$ derived from the stellar mass and radius presented in Sect.~\ref{sec:star}. The transit parameter estimates are consistent well within $\sim$\,1$\sigma$ with those derived with \texttt{allesfitter}, supporting our results.

We further carried out an independent analysis of the {\it CHEOPS} IOC LC using the {\tt pycheops}\footnote{\url{https://github.com/pmaxted/pycheops}} package (Maxted et al., in prep), which is being developed specifically for the analysis of {\it CHEOPS} data \citep[see][for more details]{lendl2020}. From this additional analysis, we obtained values (see Table~\ref{tab:allesfitter-pycheops}) fully compatible with \texttt{allesfitter} results, namely with differences lying below 1.2$\sigma$.

\begin{table}
  \centering
  \caption{Results obtained from the \texttt{pycheops} analysis of the {\it CHEOPS} IOC LC (first visit). Differences are computed with respect to the CHEOPS-only results listed in Table~\ref{tab:parameters}.}
  \label{tab:allesfitter-pycheops}
  \begin{tabular}{ccc}
    \hline
    \hline
    Parameter &  $\texttt{pycheops}$ & Difference[$\sigma$] \\   
    \hline
    \multicolumn{3}{c}{Planet c} \\
    \hline
    d$F$ [ppm] & 530$^{+67}_{-58}$ & 0.19 \\
    $W$ [h] & 2.88$\pm$0.40 & 0.50 \\
    $b$ & 0.57$^{+0.13}_{-0.22}$ & 1.17 \\
    \hline
    \multicolumn{3}{c}{Planet e}  \\ 
    \hline
    d$F$ [ppm] & 1011$^{+53}_{-54}$ & 0.81 \\
    $W$ [h] & 4.02$^{+0.14}_{-0.15}$ & 1.11 \\
    $b$ & 0.46$^{+0.09}_{-0.13}$ & 0.04  \\
    \hline
  \end{tabular}
\end{table}

We comment here in detail on the results listed in Table~\ref{tab:parameters}. The first column lists the results obtained by D20, which may be directly compared to our TESS-only results, given that both analyses use the same dataset and employ the same analysis tool. However, D20 employed an MCMC technique implemented through the \texttt{emcee} package, considered the presence of four transiting planets, and adopted $\rho_{\star,D20}=1.94\pm0.16$ g/cm$^3$ as prior. Instead, we employed the Dynamic Nested Sample technique implemented through the \texttt{dynesty} package, considered five transiting planets, and imposed a sharper Gaussian prior on the stellar density of $\rho_{\star}=1.82\pm0.12$ g/cm$^3$. Furthermore, we assumed normal Gaussian priors on the LD coefficients (see Section~\ref{sec:analysis}), while D20 set uniform priors on the LD coefficients. Despite the differences, all the TESS-only results are consistent within $\sim$1$\sigma$, but our results have in general smaller uncertainties, most likely because of the tighter priors on the stellar density and LD coefficients.

The comparison of the results obtained considering {\it TESS} and {\it CHEOPS} separately is important in order to assess the photometric precision of {\it CHEOPS} in comparison to that of {\it TESS}. The relative uncertainties $\delta$ on the transit depth $\mathrm{d}F$ are reported in Table~\ref{tab:photPrecision}. The values of $\delta$ are a function of the telescope effective diameter, the number of observed transits, the exposure time, and the transit depth. As a matter of fact, on the one hand, $b$, $e$, and $\omega$ are rather poorly constrained in the {\it CHEOPS} LCs due to the availability of single transits with frequent gaps and this affects our capability of reconstructing the transit model, especially in case of shallow transits. On the other hand, {\it CHEOPS} guarantees a greater (relative to {\it TESS}) photometric precision for shallower transits than for deeper transit as CHEOPS observations break through the {\it TESS} photometric noise floor (which limits the photometric precision of {\it TESS} for shallow transit).
In the case of HD\,108236, one {\it CHEOPS} transit observation leads to approximately the same level of precision in $\delta$ as eight {\it TESS} transits for $\mathrm{d}F\sim$500 ppm, which corresponds to the detection of a mini-Neptune with $R_p\sim$2\, $R_{\oplus}$ around a solar-like star. For deeper transit signals ($\sim$1000\,ppm), one {\it CHEOPS} transit observation leads to a precision on the transit depth higher than that of two {\it TESS} transits, roughly by a factor of 1.3. Finally, for shallower transits ($\sim$250\,ppm), one {\it CHEOPS} transit observation leads to about the same precision on the transit depth obtained after about seven {\it TESS} transit observations. However, these conclusions depend on the length and location of gaps in the {\it CHEOPS} LCs.

\begin{table*}
\caption{Comparison of the photometric precision reached by {\it TESS (TE)} and {\it CHEOPS (CH)}, quantified by the relative uncertainty $\delta$ of the squared ratio between the planetary and stellar radius $\mathrm{d}F$. Results are also influenced by specific LC features, such as gaps, especially if they occur during the ingress or egress phases. Columns labelled with gap$_{ij}$ express the temporal percentage of gaps occurring between the $i^{\mathrm{th}}$ and the $j^{\mathrm{th}}$ contact in {\it CHEOPS} LCs.}
\label{tab:photPrecision}
\centering
\begin{tabular}{c c c c c c c c}
\hline\hline
\multirow{2}*{Planet} & \# Transits & $\mathrm{d}F$ & gap$_{12}$ & gap$_{23}$ & gap$_{34}$ & \multicolumn{2}{c}{Photometric error: $\delta$} [\%] \\
 & {\it TE}:{\it CH} & [ppm] & [\%] & [\%] & [\%] & TESS-only & CHEOPS-only \\
\hline                      
b & 12:1 & 250 & 47 & 67 & 0 & 9.3 & 12 \\ 
c &  8:1 & 500 & 0 & 38 & 100 & 5.8 & 5.4 \\
d &  4:1 & 750 & 100 & 54 & 56 & 4.8 & 7.4 \\
e &  2:1 & 1000 & 94 & 61 & 0 & 5.2 & 4.1 \\
f &  2:1 & 450 & 16 & 43 & 100 & 16 & 4.8 \\
\hline
\end{tabular}
\end{table*}

The comparison between the TESS+CHEOPS approach and the analyses performed considering {\it TESS} and {\it CHEOPS} LCs separately is not straightforward. One may generally expect that the results coming from the combined analysis fall somewhere in the middle of the range defined by the results of the separate analyses, but this is not always the case. The main reason is that the CHEOPS-only analysis is not always as robust as the TESS-only analysis. In fact, in the former we consider just one transit per planet, with the photometric signal of {\it CHEOPS} LCs that is affected by frequent gaps. In particular, we are missing the ingress and/or egress phase of each transit partially or totally (see Table~\ref{tab:photPrecision}). These are the most delicate phases of the transit as they best constrain the impact parameter $b$ and hence the orbital inclination $i_p$. The loose constraint on $b$ (due to the gaps in the LCs) and the treatment of $e$ and $\omega$ as free parameters (despite only photometric data are available) increase the degrees of freedom and hence the degeneracies within the multi-parametric transit model. This justifies the discrepancies among a few parameter values, like $\mathrm{d}F$ and $i_p$ of HD\,108236 b or $i_p$ of HD\,108236 d, involving the CHEOPS-only approach.

It is in the combined analysis where we can account for the numerous transits from {\it TESS} (which give much more indications about the transit shape, decreasing the degrees of freedom of the fitted model) and, simultaneously, on the exquisite {\it CHEOPS} photometry (which allows the refinement of all the fitted parameters as the robustness of the transit shape is guaranteed by {\it TESS}). In particular, Table~\ref{tab:RpPrecision} shows the relative uncertainties on planetary radii derived from the TESS-only approach compared with those coming from the TESS+CHEOPS approach ($e$ free, so that the comparison is homogeneous). The improvement on the radii precision can be quantified by factors ranging from 1.2 for HD\,108236\,d up to 2.6 for HD\,108236\,f.

\begin{table}
\caption{Relative uncertainty on planetary radii $\frac{\sigma R_p}{R_p}$ as computed from the TESS-only and TESS+CHEOPS ($e$ free) approaches. As the {\it TESS} extended mission will re-observe HD\,108236 again in April 2021 during one sector of observations, we also predict the expected $\frac{\sigma R_p}{R_p}$ once the new {\it TESS} data will be combined to the present-day ones. The present day contribution of {\it CHEOPS} data still guarantees a better precision on the radii of planets b, e, and f, while the predicted precision is comparable for planets c and d.}
\label{tab:RpPrecision}
\centering
\begin{tabular}{c c c c}
\hline\hline
\multirow{3}*{Planet} & \multicolumn{3}{c}{$\frac{\sigma R_p}{R_p}$ [\%]} \\
 & \multicolumn{2}{c}{TESS-only} & TESS+CHEOPS \\
 & This work & April 2021 & This work \\
\hline                      
b & 4.8 & 3.9 & 3.3 \\ 
c & 3.1 & 2.5 & 2.4 \\
d & 2.6 & 2.1 & 2.2 \\
e & 2.8 & 2.3 & 1.7 \\
f & 8.7 & 7.1 & 3.3 \\
\hline
\end{tabular}
\end{table}

This system will be observed again by {\it TESS} in April 2021 during one sector of observations. Considering that the present day TESS-only results are based upon two sectors of observations, the number of data points (hence of transits measurements) are expected to be enhanced by a factor of 1.5 in April 2021. As a result, once we can account on the entire set of {\it TESS} data, the uncertainties on planetary radii are expected to be reduced by a factor of $\sqrt{1.5}\approx1.2$. As a consequence, the predicted uncertainties on planetary radii coming from TESS-only data will be comparable with those presented here for planets c and d, while the contribution of the current {\it CHEOPS} observations will still guarantee a better precision on the radii of planets b, e, and f (see Table~\ref{tab:RpPrecision}).

As mentioned earlier, we performed the TESS+CHEOPS analysis considering two cases, one leaving $e$ as a free parameter and one setting $e\lesssim0.1$ as a prior condition. On the basis of the outcome of the stability analysis, we take as reference results those obtained with $e$ constrained to be smaller than 0.1 (last column of Table~\ref{tab:parameters}). These results are consistent with those of D20, but are more precise, especially in terms of transit depth and ephemerides. This is not surprising because, despite adding just one transit per planet, {\it CHEOPS} has a larger aperture than {\it TESS} and the {\it TESS} data span $\sim$2 months, while the addition of the {\it CHEOPS} transits increases the temporal coverage to $\sim$13 months. 

To give an idea of the refinement obtained on the transit ephemerides with the results presented here, for each planet in the system we computed the precision on the transit timing for the transit closest to the middle of the next {\it CHEOPS} observing window (1 May 2021, according to the {\it CHEOPS} Feasibility Checker\footnote{\url{https://www.cosmos.esa.int/web/cheops-guest-observers-programme/scheduling-feasibility-checker}}) and compare them with those obtained by considering the ephemerides given by D20. This comparison is presented in Table~\ref{tab:ephemerides}. The longer temporal baseline has led to a significant reduction of the uncertainties on the transit times, decreasing it from over an hour to a few minutes in all cases. 

\begin{table*}
\caption{Expected transit timings and their precision for the transit occurring closest to 1 May 2021, as computed considering $T_0$ and $P$ values given by \citet{Daylan2020} and obtained from our analysis (i.e. TESS+CHEOPS with prior constraint of $e\lesssim0.1$). The improvement is due to the longer temporal baseline of the data we used in our work. The last two columns give the timescales starting from the last {\it CHEOPS} observations after which the uncertainty on $T_0$ is expected to be $\sim$\,30\,min (drift$_{30}$) and comparable to the transit duration (drift$_{W}$). $T_0$ values are given as $T_0-$2\,450\,000.}
\label{tab:ephemerides}
\centering
\begin{tabular}{l c c c c c c}
\hline\hline
 \multirow{2}*{Planet} & \multicolumn{2}{c}{\citet{Daylan2020}} & \multicolumn{4}{c}{Our work} \\
  & $T_0$ [BJD] & $\sigma T_0$ [min] & $T_0$ [BJD] & $\sigma T_0$ [min] & drift$_{30}$ [yr] & drift$_{W}$ [yr] \\
\hline
  b & 9334.9540 & 132 & 9335.09891 & 7.3 & 6.6 & 34 \\
  c & 9335.4500 & 103 & 9335.41958 & 6.5 & 6.9 & 42 \\
  d & 9336.8165 & 81  & 9336.82421 & 6.1 & 8.0 & 66 \\
  e & 9331.0523 & 115 & 9330.98999 & 3.9 & 12.4 & 108 \\
  f & \multicolumn{2}{c}{not discovered} & 9325.01948 & 9.6 & 4.1 & 29 \\
\hline
\end{tabular}
\end{table*}

For each planet, we also computed the timescale after which the 1$\sigma$-uncertainty on the transit timing becomes comparable to the transit duration, to establish the epoch when we would likely miss the full transit according to the present-day ephemerides. Starting from {\it CHEOPS} last observations, the reference timescales for the ephemerides' drifts vary from $\sim$29 years for planet f (358 orbits) up to $\sim$108 years for planet e (2013 orbits). \citet{dragomir20} evaluated that $\sim$\,98\% of {\it TESS} target stars re-observed by a follow-up mission nine months after {\it TESS} observations keep their ephemerides fresh (that is the uncertainty on $T_0$ is lower than 30 min) for at least two years. In our case {\it CHEOPS} observations occur $\sim$\,1 year after {\it TESS} observations, therefore our results may be comparable with the ephemerides deterioration estimated by \citet{dragomir20}. Similarly to what described before, we computed the time to be elapsed from {\it CHEOPS} last observations such that the error on $T_0$ becomes greater than 30 min, and we found that the ephemerides' drifts vary from $\sim$\,4 to $\sim$\,12 years (see Table~\ref{tab:ephemerides}), which is consistent with the predictions of \citet{dragomir20}.

\subsection{Dynamical analysis}\label{ssec:dynamical_analysis}

When determining orbital solutions in multi-planet systems it is important to verify the dynamical stability of the fitted systems, to ensure that the fits are physically plausible. Here, we describe the stability tests we performed for the fits presented in Section~\ref{sec:comparative_analysis}.

To test the stability of these fits, we used the MEGNO (Mean Exponential Growth factor of Nearby Orbits) functionality of the \textsc{Rebound} $N$-body package \citep{rein2012,rein2016}. MEGNO is a chaos indicator that can, in short-duration integrations, reveal chaotic behaviour that can, on longer timescales, lead to instability \citep{cincotta2003}. The MEGNO indicator $\Gamma$\,$\sim$\,2 for non-chaotic orbits, while $\Gamma$ diverges with time for chaotic orbits. We calculated $\Gamma$ for all the orbits in the TESS+CHEOPS ($e$ free) posteriors summarised in Table~\ref{tab:parameters}. The stellar mass was taken to be $0.869\mathrm{\,M}_\odot$, and the planet masses were taken from the atmospheric evolution analysis described in Section~\ref{sec:atmospheric_analysis} (see Table~\ref{tab:M5planets}). As we lack the full three-dimensional geometry of the orbits, orbital coplanarity was enforced. The systems were integrated with the \textsc{whfast} integrator \citep{rein2015} with a stepsize of $0.1$\,d, for $10^5$\,d or until the collision of two bodies or the ejection of a planet. We take $0.5$\,AU as our ejection radius as a somehow arbitrary, but reasonable trade-off accounting for instability; in fact this value implies an already significant change in orbital elements of more than three times the orbital separation for at least one planet. This scenario would produce such an instability that will eject the planet or lead to a collision, without needing to integrate the systems until a planet is physically lost from the system at much larger distances of $\sim$\,$10^5$\,AU or until the collision takes place. We classified systems as unstable if the MEGNO $\Gamma>3$, or if a collision or ejection occurred, and as stable otherwise.

In this way, we tested the stability of 54\,684 draws from the TESS+CHEOPS ($e$ free) posteriors. Only three had $\Gamma<3$, all others either having a higher MEGNO value, indicative of orbital chaos, or losing a planet within the $10^5$\,d integration duration. This lack of stability can be attributed to the lack of a constraint being placed on the orbital eccentricity in the LC fitting, resulting in fairly high eccentricities being assigned to the planets (particularly the outer two).

Noting this, and that D20 found that stable fits to the {\it TESS} LCs had eccentricities $\lesssim0.1$, we re-fit the LCs, this time imposing $|\sqrt{e}\cos\omega|,|\sqrt{e}\sin\omega|<0.3$, and again tested the stability of the new orbital solutions. The restriction on eccentricities significantly increased the number of stable configurations, with 21\,460 out of 57\,323 systems drawn from the posterior having $\Gamma<3$.  These runs were then directly integrated for 10\,Myr to verify their stability: 7\,829 of the systems survived for this time. For these systems, the median eccentricities of planets b to f are respectively $0.0286$, $0.0407$, $0.0200$, $0.0156$, and $0.0234$. These results also agree with what is found by \citet{vanEylen2015} and \citet{vanEylen2019}, who show that transiting multi-planet systems typically have very low (though not always zero) eccentricities.

\subsection{Planetary mass constraints through atmospheric evolution}\label{sec:atmospheric_analysis}

As reported by D20, only a few RV measurements are currently available for this system, thus D20 estimated the planetary masses employing the mass-radius probabilistic model of \citet{chen17}, obtaining 5$\pm$2, 7$\pm$2, 10$\pm$2, and 13$\pm$2\,$M_{\oplus}$ for planets HD\,108236\,b, HD\,108236\,c, HD\,108236\,d, and HD\,108236\,e, respectively. We also estimate the planetary masses, but following a different approach, namely employing constraints provided by the system parameters and the range of possible atmospheric evolutionary tracks realising the measured planetary radii. To this end, we use the algorithm described by \citet{kubyshkina19a,kubyshkina19b}, but employ it in a slightly modified way, as described below.

\subsubsection{Model}

The framework mixes three ingredients: a model of the stellar high-energy flux (X-ray and extreme ultra-violet radiation; hereafter XUV) evolution, a model relating planetary parameters and atmospheric mass, and a model computing atmospheric mass-loss rates. For late-type stars, the stellar XUV flux out of the saturation regime depends on stellar mass and rotation period \citep[see e.g.][]{vilhu84,wright11}, the latter of which is time-dependent. To account for the different rotation histories of the host stars, the framework models the rotation period $P_{\mathrm{rot}}(\tau)$ as a power law in age $\tau$, normalised such that the computed rotation period at the present age $P_{\mathrm{rot}}(t_{\star})$ is consistent with the now measured rotation period. Following \citet{kubyshkina19a},
\begin{equation}
    P_{\mathrm{rot}}(\tau) =
    \begin{cases}
    P_{\mathrm{rot}}(t_{\star})\left(\frac{\tau}{t_{\star}} \right)^{0.566} & \tau\geq2\: \mathrm{Gyr} \\
    P_{\mathrm{rot}}(t_{\star})\left(\frac{2}{t_{\star}} \right)^{0.566} \left(\frac{\tau}{2} \right)^x & \tau<2\: \mathrm{Gyr}
    \end{cases}
    \,,
    \label{eq:Prot}
\end{equation}
where ages are expressed in Gyr and rotation periods in days. Equation~(\ref{eq:Prot}) mimics the power-law relation given by \citet{mamajek08} for $\tau\geq2$\,Gyr, while it adjusts the power law through the $x$ exponent in the young star regime as different rotators follow different evolutionary paths.

The stellar XUV luminosity is then derived from the rotation period using scaling relations \citep{pizzolato03,sanzForcada11,wright11,mcdonald19}. To account for the evolution of the stellar bolometric luminosity, and hence the planetary equilibrium temperature ($T_{\mathrm{eq}}$), the framework uses the MIST grid. For estimating the planetary atmospheric mass fraction as a function of mass, radius, and $T_{\mathrm{eq}}$, the original framework employs models described by \citet{stokl15} and \citet{johnstone15b}. The planetary mass-loss rates are extracted from a large grid that has been constructed using the hydrodynamic model described in \citet{kubyshkina18b}. Starting at 5 Myr (the assumed age of the dispersal of the protoplanetary disk), the framework extracts the mass-loss rate from the grid at each time step, employing the stellar flux and system parameters, and uses it to update the atmospheric mass fraction and planetary radius. This procedure is then repeated until the age of the system is reached or the planetary atmosphere is completely escaped. 

The standard free parameters of the framework are the $x$ index of the power law, controlling the stellar rotation period (a proxy for the stellar XUV emission) within the first 2 Gyr, and the initial planetary radius (i.e. the initial atmospheric mass fraction at the time of the dispersal of the protoplanetary disk). The free parameters are constrained by implementing the atmospheric evolution algorithm in a Bayesian framework, employing the MCMC tool developed by \citet{cubillos17}. In short, the framework takes the observed system parameters (essentially stellar $P_{\mathrm{rot}}$, $t_{\star}$, and $M_{\star}$, besides planetary masses and orbital semi-major axes) with their uncertainties as input (i.e. priors). Then it computes millions of planetary evolutionary tracks varying the input parameters according to the shape of the prior distributions, and varying the free parameters within ranges given by the user, fitting the observed planetary radius at the age of the system, further accounting for their uncertainties. The results are posterior distributions of the free parameters, which are the rotation period of the star when it was young and the initial atmospheric mass fraction of the considered planets.

Here we reversed the usual way the tool works, that is we asked the framework to provide posteriors for the planetary masses by fixing the past rotation rate of the host star (hence its activity level) to the statistically most likely value. In particular, we imposed a normal prior on stellar age, mass, and planetary semi-major axes, according to the values derived from the host star characterisation and from the LC analysis. To describe the evolution of the activity of the host star, we fixed the stellar rotation period at an age of 150 Myr to $P_{\mathrm{rot,150}}=5.23$ days. This is the median value of the rotation period distribution given by \citet{johnstone15a}, which we inferred considering the subset of stars having masses in the 0.2~$M_{\odot}$-width interval centred around our nominal stellar mass value of $M_{\star}=0.869\, M_{\odot}$. Therefore, we run the framework considering as free parameters the present day rotational period of the star, the initial atmospheric mass fractions, and the planetary masses. The present-day stellar rotation period is formally unknown (no rotational modulation signal detected in the {\it TESS} LCs; D20), but, having fixed $P_{\mathrm{rot}}$ at the age of 150\,Myr, it is constrained by the imposed power law on the evolution of $P_{\mathrm{rot}}$. 

\subsubsection{Results}

\begin{figure*}
\hspace{-3cm}
\resizebox{22cm}{!}{\includegraphics[width=\columnwidth]{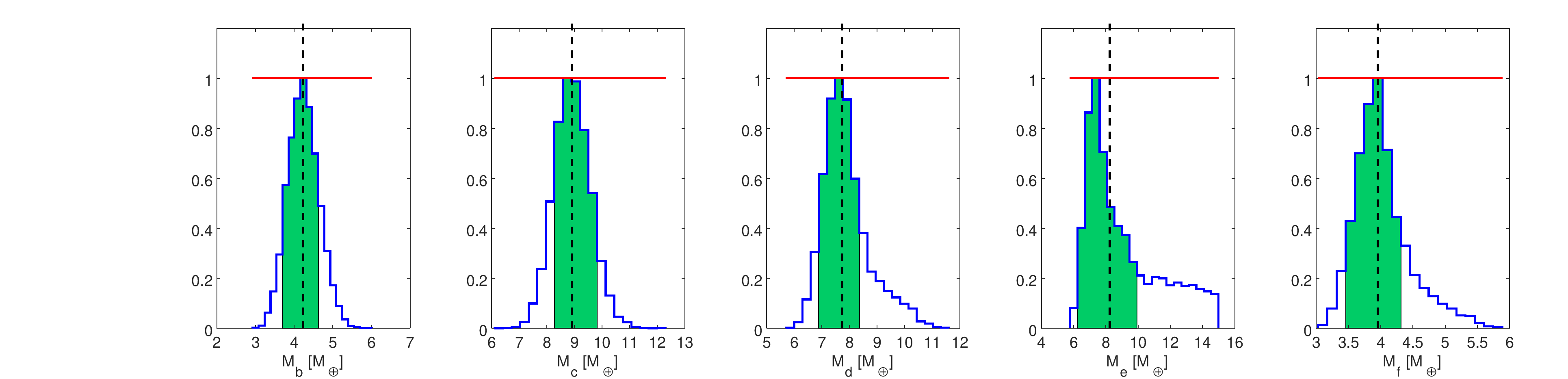}}
  \caption{Posterior distributions for the masses $M_{\mathrm{p}}$ of the planets composing the HD\,108236 system. In each panel, the horizontal red line represents the flat prior imposed on $M_{\mathrm{p}}$, implying that $M_{\mathrm{p}}$ was set as a free parameter. The dashed black line marks the median of the distribution, while the green area shows the 68\% highest posterior density credible interval.}
  \label{fig:M5planets}
\end{figure*}

Figure~\ref{fig:M5planets} shows the posterior distributions of the planetary masses we obtained at the end of the run, while Table~\ref{tab:M5planets} compares our results to those in D20 according to the probabilistic mass-radius relation of \citet{chen17}. Our results suggest that HD\,108236\,b and HD\,108236\,c have an Earth-like density, while the three outer planets should host a low mean molecular weight atmosphere. In general, there is a good agreement between our results and those obtained by D20, except for the mass of HD\,108236\,e, where our results indicate a significantly lighter planet, though our posterior distribution is skewed such that the 13$\pm$2\,$M_{\oplus}$ value proposed by D20 falls inside our upper error bar. However, we agree with D20 on the overall conclusion that this planet is a mini-Neptune. 

However, these results have to be taken with caution because they are based on models and may be affected by our assumptions. On the one hand, our results are significantly constrained by the fact that the framework models all planets in the system simultaneously, trying to find the solution that best fits them all. On the other hand, one of the main assumptions of the framework is that the analysed planets have (or had) a hydrogen-dominated atmosphere and that the planetary orbital separation does not change after the dispersal of the protoplanetary disk. Although there is reason to believe the first assumption is adequate for the vast majority of planets \citep{owen2020b}, the second assumption is most likely true for tightly packed systems with orbital resonances \citep{kubyshkina19b}, but this is not the case of HD\,108236. Indeed, both the D20 and our stability analyses (Section~\ref{ssec:dynamical_analysis}) suggest that the planets may have exchanged orbital momentum throughout their evolution, which would have also changed the planetary orbits, in turn affecting the evolution of the planetary atmospheres. From the system parameters listed in Tables~\ref{tab:stellarParam} and \ref{tab:parameters}, and from the planetary masses obtained through the atmospheric evolution calculations (Table~\ref{tab:M5planets}), we estimated the semi-amplitude ($K$) of the RV curve expected for each planet. The resulting values are listed in Table~\ref{tab:M5planets}.

\begin{table}
\caption{Estimates of planetary masses, densities, and RV semi-amplitudes according to our atmospheric evolution modelling framework and system parameters. Planetary masses and density values are compared with the predictions reported by D20, who used the probabilistic mass-radius relation of \citet{chen17}.}
\label{tab:M5planets}
\centering
\begin{tabular}{l l l l}
\hline\hline                 
\multicolumn{2}{c}{Parameters} & This work & D20 \\
\hline
\multirow{2}*{HD\,108236\,b} & $M_p$ [$M_{\mathrm{\oplus}}$] & $4.23_{-0.39}^{+0.41}$ & $5\pm2$ \\
                       & $\rho_p$ [$\rho_{\mathrm{\oplus}}$] & $1.00\pm0.10$ & $1.25\pm0.40$ \\
                       & $K$ [m\,s$^{-1}$]                   & $1.91\pm0.19$ & \\
\hline
\multirow{2}*{HD\,108236\,c} & $M_p$ [$M_{\mathrm{\oplus}}$] & $8.90_{-0.64}^{+0.67}$ & $7\pm2$ \\
                       & $\rho_p$ [$\rho_{\mathrm{\oplus}}$] & $1.00\pm0.08$ & $0.79\pm0.29$ \\
                       & $K$ [m\,s$^{-1}$]                   & $3.41\pm0.28$ & \\
\hline
\multirow{2}*{HD\,108236\,d} & $M_p$ [$M_{\mathrm{\oplus}}$] & $7.75_{-0.62}^{+0.91}$ & $10\pm2$ \\
                       & $\rho_p$ [$\rho_{\mathrm{\oplus}}$] & $0.47_{-0.08}^{+0.12}$ & $0.50\pm0.20$ \\
                       & $K$ [m\,s$^{-1}$]                   & $2.25\pm0.24$ & \\
\hline
\multirow{2}*{HD\,108236\,e} & $M_p$ [$M_{\mathrm{\oplus}}$] & $8.2_{-1.2}^{+3.8}$ & $13\pm2$ \\
                       & $\rho_p$ [$\rho_{\mathrm{\oplus}}$] & $0.28_{-0.15}^{+0.46}$ & $0.43\pm0.16$ \\
                       & $K$ [m\,s$^{-1}$]                   & $2.14\pm0.66$ & \\
\hline
\multirow{2}*{HD\,108236\,f} & $M_p$ [$M_{\mathrm{\oplus}}$] & $3.95_{-0.32}^{+0.46}$ & \multirow{2}*{not discovered} \\
                       & $\rho_p$ [$\rho_{\mathrm{\oplus}}$] & $0.48_{-0.08}^{+0.12}$ &  \\
                       & $K$ [m\,s$^{-1}$]                   & $0.90\pm0.10$ & \\
\hline
\end{tabular}
\end{table}

\section{Conclusions}\label{sec:conclusions}

We have presented the combined analysis of {\it TESS} and {\it CHEOPS} LCs of the HD\,108236 multi-planet system, along with the characterisation of the host star. Our spectroscopic analysis confirmed that HD\,108236 is a Sun-like star with $T_{\mathrm{eff}}=5660\pm61$\,K and $\log{g}=4.49\pm0.11$, though it is slightly metal poor with a metallicity of [Fe/H]$=-0.28\pm0.04$\,dex. We estimated the stellar radius to be $R_{\star}=0.877\pm0.008\,R_{\odot}$, and determined robust values for stellar mass and age by employing two sets of stellar isochrones and tracks, obtaining $M_{\star}=0.869_{-0.048}^{+0.050}\,M_{\odot}$ and $t_{\star}=6.7_{-5.1}^{+4.0}$\,Gyr.

Four planets were identified in the {\it TESS} LCs of HD\,108236 by D20. We complemented this analysis by adding three {\it CHEOPS} datasets, which were supposed to contain one further transit for each of these planets. Inspecting one of the datasets, we serendipitously discovered an unexpected transit-like feature with a depth of $\sim$400 ppm that we ascribed to a fifth planet, HD\,108236\,f. We derived the Bayesian evidence for a four- and five-planet model, finding that the latter was significantly favoured. The combined {\it TESS} and {\it CHEOPS} LC analysis led us to derive a likely period for HD\,108236\,f of $P_{\mathrm{f}}=29.54$\,days.
Within the context of searching for additional planets in the system, we also investigated the presence in the {\it CHEOPS} LCs of the tentative candidate with a period of 10.9113 days suggested by D20, but we could not find strong evidence supporting it.

Within the scenario comprising five planets, the combined analysis of the {\it TESS} and {\it CHEOPS} LCs resulted in system parameters in agreement with those of D20 for planets HD\,108236\,b to HD\,108236\,e. However, our results are more precise in terms of transit depths (due to the high quality of the {\it CHEOPS} photometry), planetary radii (due to the improved measured transit depths and stellar radius), and ephemerides (due to the longer baseline covered by the data). We further ran a dynamical stability analysis of the system, finding that stability is maximised when the eccentricities of all planets are smaller than about 0.1. Comparing the results obtained from the analysis of the {\it TESS} and {\it CHEOPS} LCs separately, we found that, for a $V$\,$\sim$\,9\,mag solar-like star considering a transit signal with a depth of $\sim$500\,ppm (i.e. a $\sim$2 $R_{\oplus}$ mini-Neptune transiting a solar-like star), one {\it CHEOPS} transit returns data with a quality comparable to that obtained from about eight {\it TESS} transits.

The refined ephemerides we obtained from the LC analysis will allow more accurate planning of follow-up observations of this system. HD\,108236 will be visible with {\it CHEOPS} again, with an efficiency higher than 50\%, between March and June 2021, at which point the planetary transit timings will be predictable with uncertainties of just a few minutes. Instead, by employing the $T_0$ and $P$ values from D20, based on just $\sim$\,2~months of {\it TESS} observations, the uncertainties on the transit timings would have been of the order of a couple of hours. We also evaluated the timescales after which the uncertainties on the transit timings become greater than 30 minutes, following the criterion in \citet{dragomir20}. We concluded that our ephemerides remain fresh on timescales ranging between $\sim$\,4~years (planet f) and $\sim$\,12 years (planet e) from the last {\it CHEOPS} observations.

{\it TESS} will observe HD\,108236 again during the extended mission in April 2021 within one sector of observations. If combined with the already available observations, and assuming {\it TESS} will collect the exact same amount of transits for each planet per sector, the entire {\it TESS} dataset will help reduce the uncertainties on the planetary radii by a factor of $\sim$\,1.2. This would imply reaching the same level of precision as our TESS+CHEOPS approach for planets c and d; however, the results presented here with the contribution of {\it CHEOPS} data would still guarantee a better precision on the radii of planets b, e, and f by at least a factor of $\sim$\,1.2. Future {\it CHEOPS} observations will therefore lead to further improvements in the measurement of the planetary radii, even more so compared to what was obtained through the {\it TESS} extended mission data.
Furthermore, future {\it CHEOPS} data on this system will be important for looking for and measuring the TTVs that provide key information regarding the system dynamics as well as a measurement of the planetary masses, independent of what given by the RV method.

We finally used the derived system parameters and a planetary atmospheric evolutionary framework to constrain the planetary masses, finding that HD\,108236\,b and HD\,108236\,c should have an Earth-like density, while the outer planets are likely to host a low mean molecular weight atmosphere. The brightness and age of the host star make this an ideal target for measuring planetary masses using high-precision RV measurements. The expected semi-amplitudes of the RV variations are within reach of instruments such as HARPS \citep{Mayor2003} and ESPRESSO \citep{Pepe2020} spectrographs. This will enable us to check the impact of the modelling assumptions and to perform more detailed planetary atmospheric evolution modelling, aimed at deriving the past and future evolution of the system.

\begin{acknowledgements}
We are extremely grateful to the anonymous referee for the very thorough comments, which definitely improved the quality of this manuscript.
CHEOPS is an ESA mission in partnership with Switzerland with important contributions to the payload and the ground segment from Austria, Belgium, France, Germany, Hungary, Italy, Portugal, Spain, Sweden, and the United Kingdom.
The Swiss participation to CHEOPS has been supported by the Swiss Space Office (SSO) in the framework of the Prodex Programme and the Activités Nationales Complémentaires (ANC), the Universities of Bern and Geneva as well as well as of the NCCR PlanetS and the Swiss National Science Foundation. 
Based on observations made with the ESO 3.6\,m telescope at the La Silla Observatory under program ID\,1102.C-0923.
M.Le acknowledges support from the Austrian Research Promotion Agency (FFG) under project 859724 “GRAPPA”. 
A.De and D.Eh acknowledge support from the European Research Council (ERC) under the European Union’s Horizon 2020 research and innovation programme (project Four Aces; grant agreement No 724427).
M.J.Ho acknowledges the support of the Swiss National Fund under grant 200020\_172746.
B.-O.De acknowledges support from the Swiss National Science Foundation (PP00P2-190080).
The Spanish scientific participation in CHEOPS has been supported by the Spanish Ministry of Science and Innovation and the European Regional Development Fund through grants ESP2016-80435-C2-1-R, ESP2016-80435-C2-2-R, ESP2017-87676-C5-1-R, PGC2018-098153-B-C31, PGC2018-098153-B-C33, and MDM-2017-0737 Unidad de Excelencia María de Maeztu–Centro de Astrobiología (INTA-CSIC), as well as by the Generalitat
de Catalunya/CERCA programme.
The MOC activities have been supported by the ESA contract No. 4000124370. 
This work was supported by Fundação para a Ciência e a Tecnologia (FCT) through national funds and by Fundo Europeu de Desenvolvimento Regional (FEDER) via COMPETE2020 - Programa Operacional Competitividade e Internacionalização through the research grants: UID/FIS/04434/2019; UIDB/04434/2020; UIDP/04434/2020; PTDC/FIS-AST/32113/2017 \& POCI-01-0145-FEDER-032113; PTDC/FIS-AST/28953/2017 \& POCI-01-0145-FEDER-028953;
PTDC/FIS-AST/28987/2017 \& POCI-01-0145-FEDER-028987. 
S.C.C.Ba and S.G.So acknowledge support from FCT through FCT contracts nr. IF/01312/2014/CP1215/CT0004, IF/00028/2014/CP1215/CT0002. 
O.D.S.De is supported in the form of work contract (DL 57/2016/CP1364/CT0004) funded by national funds through Fundação para a Ciência e Tecnologia (FCT). 
The Belgian participation to CHEOPS has been supported by the Belgian Federal Science Policy Office (BELSPO) in the framework of the PRODEX Program, and by the University of Liège through an ARC grant for Concerted Research Actions financed by the Wallonia-Brussels Federation. M.Gi is F.R.S.-FNRS Senior Research Associate. 
S.Sa has received funding from the European Research Council (ERC) under the European Union’s Horizon 2020 research and innovation programme (grant agreement No 833925, project STAREX).
Gy.M.Sz acknowledges funding from the Hungarian National Research, Development and Innovation Office (NKFIH) grant GINOP-2.3.2-15-2016-00003 and K-119517. 
For Italy, CHEOPS actvities have been supported by the Italian Space Agency, under the programs: ASI-INAF n. 2013-016-R.0 and ASI-INAF n. 2019-29-HH.0.
L.Bo, G.Pi, I.Pa, G.Sc, and V.Na acknowledge the funding support from Italian Space Agency (ASI) regulated by ``Accordo ASI-INAF n. 2013-016-R.0 del 9 luglio 2013 e integrazione del 9 luglio 2015''.
G.La acknowledges support by CARIPARO Foundation, according to the agreement CARIPARO-Universit{\`a} degli Studi di Padova (Pratica n. 2018/0098).
A.Mu acknowledges support from the Swedish National Space Agency (grant 120/19~C). The dynamical simulations were enabled by resources provided by the Swedish National Infrastructure for Computing (SNIC) at Lunarc partially funded by the Swedish Research Council through grant agreement no. 2016-07213. Simulations in this paper made use of the REBOUND code which is freely available at \url{http://github.com/hannorein/rebound}.
S.Ho acknowledges CNES funding through the grant 837319.
K.G.I. is the ESA CHEOPS Project Scientist and is responsible for the ESA CHEOPS Guest Observers Programme. She does not participate in, or contribute to, the definition of the Guaranteed Time Programme of the CHEOPS mission through which observations described in this paper have been taken, nor to any aspect of target selection for the programme.
X.Bo, S.Ch, D.Ga, M.Fr, and J.La acknowledge their roles as ESA-appointed CHEOPS science team members.
A.Bo acknowledges B. Akinsanmi, G. Bruno, M. Günther, R. Luque, F.J. Pozuelos Romero, and L.M. Serrano for the very fruitful discussions.
We acknowledge T. Daylan for his help in planning the {\it CHEOPS} observations.
\end{acknowledgements}

\bibliography{biblio}
\bibliographystyle{aa}

\begin{appendix}
\section{Initial priors of the LC analyses.}

\begin{table}[h!]
\caption{Initial priors used in our fits. $\mathcal{N}(\mu_0,\sigma_0)$ denotes a Normal (Gaussian) prior with mean $\mu_0$ and standard deviation $\sigma_0$, while $\mathcal{U}(a_0, b_0)$ denotes a uniform prior, whose bounds are $a_0$ and $b_0$. $\rho_{\star}$ is expressed in g/cm$^3$, $T_0$ in BJD, $P$ in days. All other quantities are dimensionless.}
\label{tab:priors}
\centering
\begin{tabular}{l c}
\hline\hline            
 Parameter & Prior \\
 $\rho_{\star}$ & $\mathcal{N}(1.82,0.12)$ \\
 $q_{1,{\mathrm{TESS}}}$ & $\mathcal{N}(0.34,0.05)$ \\
 $q_{2,{\mathrm{TESS}}}$ & $\mathcal{N}(0.27,0.05)$ \\
 $q_{1,{\mathrm{CHEOPS}}}$ & $\mathcal{N}(0.46,0.05)$ \\
 $q_{2,{\mathrm{CHEOPS}}}$ & $\mathcal{N}(0.32,0.05)$ \\
\hline
\multicolumn{2}{c}{Planet b} \\
$\frac{R_p}{R_{\star}}$ & $\mathcal{N}(0.016,0.005)$ \\
$\frac{R_p+R_{\star}}{a}$ & $\mathcal{N}(0.091,0.008)$ \\
$\cos{i}$ & $\mathcal{U}(0.,0.25)$ \\
$T_0$ & $\mathcal{N}(2458598.68,0.02)$ \\
$P$ & $\mathcal{N}(3.7955,0.0043)$ \\
$\sqrt{e}\cos{\omega}$ & $\mathcal{U}(-0.7,0.7)$ \\
$\sqrt{e}\sin{\omega}$ & $\mathcal{U}(-0.7,0.7)$ \\
\hline
\multicolumn{2}{c}{Planet c} \\
$\frac{R_p}{R_{\star}}$ & $\mathcal{N}(0.021,0.003)$ \\
$\frac{R_p+R_{\star}}{a}$ & $\mathcal{N}(0.065,0.006)$ \\
$\cos{i}$ & $\mathcal{U}(0.,0.25)$ \\
$T_0$ & $\mathcal{N}(2458597.210,0.014)$ \\
$P$ & $\mathcal{N}(6.2036,0.0035)$ \\
$\sqrt{e}\cos{\omega}$ & $\mathcal{U}(-0.7,0.7)$ \\
$\sqrt{e}\sin{\omega}$ & $\mathcal{U}(-0.7,0.7)$ \\
\hline
\multicolumn{2}{c}{Planet d} \\
$\frac{R_p}{R_{\star}}$ & $\mathcal{N}(0.026,0.004)$ \\
$\frac{R_p+R_{\star}}{a}$ & $\mathcal{N}(0.039,0.004)$ \\
$\cos{i}$ & $\mathcal{U}(0.,0.25)$ \\
$T_0$ & $\mathcal{N}(2458599.688,0.011)$ \\
$P$ & $\mathcal{N}(14.1757,0.0079)$ \\
$\sqrt{e}\cos{\omega}$ & $\mathcal{U}(-0.9,0.9)$ \\
$\sqrt{e}\sin{\omega}$ & $\mathcal{U}(-0.9,0.9)$ \\
\hline
\multicolumn{2}{c}{Planet e} \\
$\frac{R_p}{R_{\star}}$ & $\mathcal{N}(0.032,0.003)$ \\
$\frac{R_p+R_{\star}}{a}$ & $\mathcal{N}(0.031,0.002)$ \\
$\cos{i}$ & $\mathcal{U}(0.,0.25)$ \\
$T_0$ & $\mathcal{N}(2458606.160,0.012)$ \\
$P$ & $\mathcal{N}(19.591,0.013)$ \\
$\sqrt{e}\cos{\omega}$ & $\mathcal{U}(-0.9,0.9)$ \\
$\sqrt{e}\sin{\omega}$ & $\mathcal{U}(-0.9,0.9)$ \\
\hline
\multicolumn{2}{c}{Planet f} \\
$\frac{R_p}{R_{\star}}$ & $\mathcal{N}(0.020,0.006)$ \\
$\frac{R_p+R_{\star}}{a}$ & $\mathcal{N}(0.024,0.004)$ \\
$\cos{i}$ & $\mathcal{U}(0.,0.25)$ \\
$T_0$ & $\mathcal{N}(2458616.034,0.058)$ \\
$P$ & $\mathcal{N}(29.542,0.016)$ \\
$\sqrt{e}\cos{\omega}$ & $\mathcal{U}(-0.9,0.9)$ \\
$\sqrt{e}\sin{\omega}$ & $\mathcal{U}(-0.9,0.9)$ \\
\hline                                   
\end{tabular}
\end{table}

\end{appendix}

\end{document}